\documentclass{scrartcl}

\DeclareOldFontCommand{\rm}{\normalfont\rmfamily}{\mathrm}
\DeclareOldFontCommand{\sf}{\normalfont\sffamily}{\mathsf}
\DeclareOldFontCommand{\tt}{\normalfont\ttfamily}{\mathtt}
\DeclareOldFontCommand{\bf}{\normalfont\bfseries}{\mathbf}
\DeclareOldFontCommand{\it}{\normalfont\itshape}{\mathit}
\DeclareOldFontCommand{\sl}{\normalfont\slshape}{\@nomath\sl}
\DeclareOldFontCommand{\sc}{\normalfont\scshape}{\@nomath\sc}
\DeclareRobustCommand*\cal{\@fontswitch\relax\mathcal}
\DeclareRobustCommand*\mit{\@fontswitch\relax\mathnormal}

\usepackage[utf8]{inputenc}
\usepackage{color}
\usepackage{type1cm}
\usepackage{graphicx}
\usepackage[usenames,svgnames]{xcolor}
\usepackage[hyphens]{url}
\usepackage[colorlinks=true,linkcolor=blue,citecolor=blue,breaklinks=true]{hyperref}

\usepackage[numbers,sort&compress]{natbib}
\usepackage{amsmath,amssymb,bm,amsthm,cases}

\def\beq{\begin{equation}}
\def\eeq{\end{equation}}
\def\<{\langle}
\def\>{\rangle}
\def\Tr{\mathrm{Tr}}
\renewcommand{\d}{\partial}

\begin{document}
\title{Prethermalization in the transverse-field Ising chain with long-range interactions}
\author{Takashi Mori \\
}
%\date{\large\today}
\date{\normalsize{\textit{RIKEN Center for Emergent Matter Science (CEMS), Wako 351-0198, Japan
}}}
\maketitle

\begin{abstract}
Nonequilibrium dynamics of an isolated quantum spin chain with long-range Ising interactions that decay as $1/r^\alpha$ ($0<\alpha<1$) with the distance $r$ is studied.
It turns out that long-range interactions give rise to a big timescale separation, which causes prethermalization for all $\alpha\in(0,1)$.
This conclusion is deduced by comparing two important timescales relevant for relaxation dynamics; one is the relaxation time of local permutation operators, which are quasi-conserved quantities in this system, and the other is the timescale of the initial relaxation due to the growth of quantum fluctuations.
We also explore the entire nonequilibrium dynamics by using the discrete truncated Wigner approximation, which is consistent with the result mentioned above. 
\end{abstract}

\section{Introduction}
\label{sec:intro}

Isolated many-body quantum systems show intriguing nonequilibrim dynamics.
If the system is initially in a pure state $|\Psi(0)\>$, the quantum state at time $t$, $|\Psi(t)\>$, remains pure and it evolves under the Schr\"odinger equation $id|\Psi(t)\>/dt=\hat{H}|\Psi(t)\>$, where $\hat{H}$ denotes the Hamiltonian and we put $\hbar=1$ throughout the paper.
Although $|\Psi(t)\>$ does not reach any stationary quantum state unless the initial state is stationary, the expectation value $\<\Psi(t)|\hat{O}|\Psi(t)\>$ of a local observable $\hat{O}$ can exhibit \textit{equilibration}~\cite{Reimann2008,Short2012,Reimann2012}, i.e., the approach to a stationary value, or even \textit{thermalization}~\cite{Neumann1929,Deutsch1991,Srednicki1994,Tasaki1998,Rigol2008,Biroli2010}, i.e., the approach to its equilibrium value $\<\hat{O}\>_{\mathrm{eq}}$ predicted by statistical mechanics.
It is due to the effect of quantum entanglement~\cite{Popescu2006,Goldstein2006}, and there is no classical counterpart\footnote
{In a classical system, properly defined macroscopic quantities can thermalize, but local quantities do not; they remain fluctuating largely.
}.

Before reaching thermal equilibrium, an isolated quantum system sometimes exhibits a two-step relaxation, i.e., $\<\Psi(t)|\hat{O}|\Psi(t)\>$ reaches a quasi-stationary value $\<\hat{O}\>_{\mathrm{pre}}$, which differs from $\<\hat{O}\>_{\mathrm{eq}}$, and stay there in a certain long timescale.
After a relaxation time $\tau_{\mathrm{rel}}$, $\<\Psi(t)|\hat{O}|\Psi(t)\>$ will eventually thermalize.
This phenomenon is referred to as \textit{prethermalization}~\cite{Berges2004}.
Prethermalization is a consequence of the existence of a big timescale separation.
That is, there will exist some \textit{quasi-conserved quantities} which are almost conserved up to the time $\tau_{\mathrm{rel}}$ that is much longer than the timescale of the initial relaxation $\tau_{\mathrm{ini}}$.
For a recent theoretical review, see Ref.~\cite{Mori2018_review}.

Prethermalization has been studied theoretically in several short-range interacting systems such as field theories~\cite{Berges2004,Berges2008,Sciolla2013}, fermion systems~\cite{Kollar2008,Moeckel2008,Eckstein2009,Schiro2010,Kollar2011,Bertini2015}, boson systems~\cite{Kollath2007,Kitagawa2010,Kitagawa2011,Kaminishi2015,Kaminishi2018}, and periodically driven systems~\cite{Bukov2015,Kuwahara2016_Floquet,Mori2016_rigorous,Abanin2017}.
Experimentally, prethermalization was observed in one-dimensional Bose gas after a coherent splitting~\cite{Gring2012,Smith2013,Langen2015}.
It has been also reported that long-range interactions decaying as $1/r^{\alpha}$ with distance $r$ give rise to prethermalization~\cite{Marcuzzi2013,Gong2013,Worm2013,Kastner2015,Smacchia2015}, where $\alpha$ satisfies $0\leq\alpha<d$ with the spatial dimension $d$. 
Dynamical properties of a long-range interacting system will significantly differ from those of a short-range interacting one.
For example, long-range interacting spin systems do not have any sharp Lieb-Robinson bound~\cite{Hastings2006,Eisert2013}, but sometimes fast propagation of perturbation is suppressed~\cite{Santos2016} and the entanglement growth is very slow~\cite{Schachenmayer2013,Buyskikh2016}.
Recent experiments with ion traps enable us to realize a spin system with tunable long-range interactions~\cite{Porras2004,Kim2009,Britton2012,Islam2013,Neyenhuis_arXiv2016}, and some intriguing dynamics has been observed~\cite{Philip2014,Smith2016}.
Thus, nonequilibrium dynamics of long-range interacting systems is interesting in its own right.

It is however very difficult to theoretically study (pre)thermalization in long-range interacting systems.
Long-time dynamics can be computed through the exact diagonalization of the Hamiltonian, but this method is limited to small system sizes.
Recent studies have calculated nonequilibrium dynamics of long-range interacting spins up to $N\simeq 100$, where $N$ denotes the number of spins, by using the time-dependent density matrix renormalization group method~\cite{Schachenmayer2013} or the variational method based on the matrix product states~\cite{Buyskikh2016}.
However, finite-size effects are very strong in long-range interacting systems, and hence we cannot fully understand their nonequilibrium dynamics by those methods.
The mean-field approximation is an efficient theoretical tool to study long-range interacting systems, and it enables us to study the dynamics accurately in the thermodynamic limit.
However, in a long-range interacting system, relaxation timescales diverge in the thermodynamic limit~\cite{Kastner2011}, and hence the long-time dynamics in the thermodynamic limit is quite different from that in a large but finite system.
Only the latter can exhibit (pre)thermalization in a long-range interacting spin system.
We must study the long-time dynamics of a large but finite system in order to fully understand the relaxation dynamics; it is a challenging problem.

This paper focuses on a long-range interacting spin chain ($d=1$).
Analytical studies on an exactly solvable model called the generalized Emch-Radin model~\cite{Emch1966,Radin1970} revealed that prethermalization is present in this model for $0<\alpha\leq 1/2$ but not for $1/2<\alpha<1$~\cite{Worm2013,Kastner2015}.
This is a remarkable result, but it has not been addressed whether the same type of prethermalization is present in more general long-range interacting spin systems.
In this paper, this problem is solved for a spin chain with long-range Ising interactions and the transverse field.
It turns out that prethermalization for $0<\alpha\leq 1/2$ found in the exactly solvable model persists in a more general case, and furthermore, it turns out that prethermalization should also occur for $1/2<\alpha<1$ in general.
The absence of prethermalization for $1/2<\alpha<1$ in the generalized Emch-Radin model~\cite{Worm2013,Kastner2015} should be understood as an exceptional result due to the special feature of an exactly solvable model. 

The analysis goes as follows.
First, we identify quasi-conserved quantities by proving that every local permutation operator (defined later) is a quasi-conserved quantity which changes its value in a timescale $\tau_{\mathrm{rel}}\propto N^{1-\alpha}$.
Next, we numerically evaluate the timescale $\tau_{\mathrm{ini}}$ of the initial stage of relaxation by solving the approximate equations of motion for spin operators with the leading-oder corrections in $1/N$.
As a result, it turns out that $\tau_{\mathrm{ini}}\propto\ln N$, and hence $\tau_\mathrm{rel}$ is much larger than $\tau_{\mathrm{ini}}$ for any value of $\alpha\in(0,1)$. 
This implies that there is a big timescale separation and prethermalization should be present for any $\alpha\in(0,1)$.
Finally, we calculate the entire nonequilibrium dynamics of the present model by using the discrete truncated Wigner approximation (DTWA), which is a recently proposed semiclassical approximation for quantum systems with discrete degrees of freedom~\cite{Schachenmayer2015,Pucci2016}.
DTWA calculations are consistent with the conclusion that prethermalization occurs for any $\alpha\in(0,1)$.
Although we focus on the transverse-field Ising model, the result in this paper is expected to be true for more general class of long-range interacting spin systems.

The remaining part of this paper is organized as follows.
In section~\ref{sec:model}, the model is explained.
In section~\ref{sec:simple}, the previously known results in two simple cases are summarized.
The main result of this paper is presented in section~\ref{sec:main}.
The proof of the existence of quasi-conserved quantities with lifetime $\tau_{\mathrm{rel}}\propto N^{1-\alpha}$ is given in subsection~\ref{sec:proof}.
In subsection~\ref{sec:timescale}, we numerically evaluate the timescale of the early stage of relaxation.
The results given in two subsections~\ref{sec:proof} and \ref{sec:timescale} lead us to the conclusion summarized above.
In section~\ref{sec:DTWA}, we show numerical results using DTWA.
It is confirmed that the DTWA calculations reproduce the results presented in Sec.~\ref{sec:main}.
Based on DTWA calculations, we argue that the logarithmic dependence of $\tau_\mathrm{ini}$ on $N$ comes from the chaoticity of the underlying classical dynamics.
We summarize our finding in section~\ref{sec:conclusion}.

\section{Model}
\label{sec:model}

We consider a spin-1/2 chain whose Hamiltonian is given by
\beq
\hat{H}=-\sum_{i<j}^NJ_{ij}\hat{\sigma}_i^z\hat{\sigma}_j^z-h_z\sum_{i=1}^N\hat{\sigma}_i^z-h_x\sum_{i=1}^N\hat{\sigma}_i^x,
\label{eq:H}
\eeq
where $\hat{\bm{\sigma}}_i$ denotes the Pauli operator at site $i$.
The first term of the right hand side of Eq.~(\ref{eq:H}) expresses long-range Ising interactions and the second and third terms represent the longitudinal and the transverse fields, respectively.
We employ the periodic boundary condition, and let us denote by $r_{ij}$ the distance between sites $i$ and $j$,
\beq
r_{ij}:=\min\{|i-j|,N-|i-j|\}.
\eeq
The Ising interaction $J_{ij}$ is given by
\beq
J_{ij}=\frac{J_{\alpha}}{N^{1-\alpha}}\frac{1}{r_{ij}^{\alpha}},
\eeq
where $\alpha$ satisfies $0\leq\alpha<1$ and $J_{\alpha}$ is chosen so that the interaction energy per spin is normalized as
\beq
\frac{J_{\alpha}}{N^{1-\alpha}}\sum_{i(\neq j)}^N\frac{1}{r_{ij}^{\alpha}}=1.
\eeq
This normalization is referred to as the Kac prescription in the literature~\cite{Kac1963,Campa_review2009}.
Because of the factor $1/N^{1-\alpha}$, $J_{\alpha}$ is finite in the thermodynamic limit.

In this paper, we focus on the initial state fully-polarized along $x$ direction,
\beq
|\Psi(0)\>=\bigotimes_{i=1}^N|+_x\>=|+_x\>\otimes|+_x\>\otimes\dots\otimes|+_x\>,
\label{eq:initial}
\eeq
where $\hat{\sigma}^x|+_x\>=|+_x\>$.
The quantum state $|\Psi(t)\>$ evolves according to the Schr\"odinger equation under the Hamiltonian $\hat{H}$ given by Eq.~(\ref{eq:H}),
\beq
i\frac{d}{dt}|\Psi(t)\>=\hat{H}|\Psi(t)\>.
\eeq
After the time evolution, $|\Psi(t)\>$ is no longer expressed as a product state like Eq.~(\ref{eq:initial}).
In the mean-field (MF) approximation, the system is assumed to be in a product state at any time $t$,
\beq
|\Psi_{\mathrm{MF}}(t)\>=\bigotimes_{i=1}^N|\psi_{\mathrm{MF}}(t)\>,
\eeq
and $|\psi_{\mathrm{MF}}(t)\>$ obeys a nonlinear differential equation determined self consistently.
For example, in the spin-1/2 Hamiltonian~(\ref{eq:H}), the MF equation reads
\beq
i\frac{d}{dt}|\psi_{\mathrm{MF}}(t)\>=\left(-\<\psi_{\mathrm{MF}}(t)|\hat{\sigma}^z|\psi_{\mathrm{MF}}(t)\>\hat{\sigma}^z-h_z\hat{\sigma}^z-h_x\hat{\sigma}^x\right)|\psi_{\mathrm{MF}}(t)\>,
\eeq
or equivalently, in the Heisenberg picture,
\beq
\left\{
\begin{split}
&\frac{d}{dt}\<\hat{\sigma}^x\>_{\mathrm{MF}}(t)=2\<\hat{\sigma}^z\>_{\mathrm{MF}}(t)\<\hat{\sigma}^y\>_{\mathrm{MF}}(t), \\
&\frac{d}{dt}\<\hat{\sigma}^y\>_{\mathrm{MF}}(t)=-2\<\hat{\sigma}^z\>_{\mathrm{MF}}(t)\<\hat{\sigma}^x\>_{\mathrm{MF}}(t)+2h\<\hat{\sigma}^z\>_{\mathrm{MF}}(t) \\
&\frac{d}{dt}\<\hat{\sigma}^z\>_{\mathrm{MF}}(t)=-2h\<\hat{\sigma}^y\>_{\mathrm{MF}}(t),
\end{split}
\right.
\label{eq:MF_Heisenberg}
\eeq
where $\<\hat{\bm{\sigma}}\>_{\mathrm{MF}}(t):=\<\psi_{\mathrm{MF}}(t)|\hat{\bm{\sigma}}|\psi_{\mathrm{MF}}(t)\>$.
In the MF approximation, the spin-spin correlations due to quantum entanglement are neglected.
The MF equation in the Heisenberg picture, Eq.~(\ref{eq:MF_Heisenberg}), can be interpreted as classical equations of motion, and hence the MF approximation is sometimes called the classical approximation.
It is shown that the MF approximation becomes exact in the limit of $N\rightarrow\infty$ in a long-range interacting system under the periodic boundary condition, see section~\ref{sec:timescale}.
When $N$ is large but finite, the MF approximation provides us a good approximation within a certain timescale diverging as $N\rightarrow\infty$.

As mentioned in Introduction, equilibration or thermalization of a local observable occurs due to the quantum entanglement.
Therefore, within the MF approximation, there is neither equilibration nor thermalization.
The central interest of the present work is in (pre)thermalization that occurs when $N$ is sufficiently large but finite.
In Sec.~\ref{sec:DTWA} we will show numerical results using a more sophisticated semiclassical approximation, i.e., the DTWA, which somehow takes into account the effect of quantum fluctuations and entanglement~\cite{Schachenmayer2015}.
We will see that the DTWA enables us to calculate the entire relaxation dynamics including both prethermalization and thermalization.

\section{Simple cases}
\label{sec:simple}

In this section, we summarize the previously known results in two simple cases, $\alpha=0$ and $h_x=0$.
Investigating these simple cases will help us to understand more general situations.

\subsection{$\alpha=0$}
\label{sec:alpha0}

The model with $\alpha=0$ is called a fully-connected model.
A remarkable property of a fully-connected model is that it has the permutation symmetry, i.e., the Hamiltonian is invariant under any permutation of spins.
This leads to a great simplification of the problem; permutation operators $\hat{P}_{ij}$ of spins $i$ and $j$ are conserved for all $i$ and $j$.
In a spin-1/2 system, the permutation operator is written in terms of Pauli operators as
\beq
\hat{P}_{ij}=\frac{1}{2}(1+\hat{\bm{\sigma}}_i\cdot\hat{\bm{\sigma}}_j).
\label{eq:permutation}
\eeq
Let us introduce the totally-symmetric subspace (TSS), which is the Hilbert subspace spanned by the states with $\hat{P}_{ij}=1$ for all $i$ and $j$.
In the spin-1/2 case, it is well known that the TSS corresponds to the subspace with the maximum total spin length, $\hat{\bm{S}}_{\mathrm{tot}}^2=(N/2)(N/2+1)$ with $\hat{\bm{S}}_{\mathrm{tot}}:=\sum_{i=1}^N\hat{\bm{\sigma}}_i/2$.
It is noted that the fully-polarized initial state is in the TSS.
Since the Hamiltonian has the permutation symmetry, the state $|\Psi(t)\>$ after a time evolution is always in the TSS.
Therefore, we can restrict our discussion into the TSS.

The relaxation dynamics for $\alpha=0$ has been recently studied in detail~\cite{Mori2017_classical}.
A crucial observation is that, within the TSS, the model with $\alpha=0$ can be considered to be a semiclassical system with the effective Planck constant $\hbar_{\mathrm{eff}}=1/N$~\cite{Sciolla2010,Sciolla2011}.
In the spin-1/2 system, any state $|\Psi(t)\>$ in the TSS is expanded as $|\Psi(t)\>=\sum_q\Psi_t(q)|q\>$, where $|q\>$ is a simultaneous eigenstate of $\hat{\bm{S}}_{\mathrm{tot}}^2$ and $\hat{S}_\mathrm{tot}^z$ with the eigenvalues $(N/2)(N/2+1)$ and $Nq$, respectively, where $Nq\in\{-N/2,-N/2+1,\dots,N/2\}$.
When $N$ is large, $q$ can be approximately regarded as a continuous variable, and we can introduce the ``canonical momentum'' conjugate to $q$ as $p:=-i\hbar_{\mathrm{eff}}\d/\d q=(-i/N)\d/\d q$.
Then, the original Schr\"odinger equation is approximated as the following Schr\"odinger equation of a single particle with the position $q$, the momentum $p$, and the effective Planck constant $\hbar_{\mathrm{eff}}=1/N$:
\beq
\frac{i}{N}\frac{\d}{\d t}\Psi_t(q)=\tilde{H}\Psi_t(q),
\eeq
with
\beq
\tilde{H}=-2q^2-2h_zq-h_x\sqrt{1-(2q)^2}\cos p
\eeq
up to the leading order in $1/N$.
The correspondence between the canonical variables $(q,p)$ and the spin variable $\bm{m}=(1/N)\sum_{i=1}^N\bm{\sigma}$ is given by
\beq
\left\{
\begin{split}
&m^x=\sqrt{1-(2q)^2}\cos p, \\
&m^y=-\sqrt{1-(2q)^2}\sin p, \\
&m^z=2q.
\end{split}
\right.
\eeq
The thermodynamic limit $N\rightarrow\infty$ corresponds to the classical approximation, in which the classical variables $(q,p)$ obey the Hamilton equations of motion,
\beq
\left\{
\begin{split}
&\frac{dq}{dt}=\frac{\d}{\d p}\tilde{H}(q,p), \\
&\frac{dp}{dt}=-\frac{\d}{\d q}\tilde{H}(q,p).
\end{split}
\right.
\label{eq:classical}
\eeq
These classical equations of motion are equivalent to the MF dynamics given by Eq.~(\ref{eq:MF_Heisenberg}).
Thus, the MF approximation is interpreted as the classical approximation, and the MF approximation becomes exact in the thermodynamic limit.

When $N$ is large but finite, the quantum dynamics is described by the truncated Wigner approximation (TWA)~\cite{Polkovnikov2010}, which is a kind of semiclassical approximations.
In this approximation, the initial state is represented as a (quasi)-probability distribution in the classical phase space, which is nothing but the Wigner function, and it obeys the \textit{purely classical} equations of motion, i.e., the Liouville equation.
Since the variables $(q,p)$ have quantum fluctuations proportional to $N^{-1/2}$ in the initial state of Eq.~(\ref{eq:initial}), the initial state is represented by a sharply localized distribution function.
The relaxation takes place through the growth of quantum fluctuations.
According to the TWA, this growth of quantum fluctuations corresponds to the spread of the distribution function over the equal-energy surface under the classical Liouville equation.

The speed of the growth of quantum fluctuations crucially depends on whether the classical dynamics is regular or chaotic~\cite{Sciolla2010,Mori2017_classical}.
When the classical dynamics is regular, quantum fluctuations grow linearly as $t/N^{1/2}$, and hence the relaxation time scales as $N^{1/2}$.
On the other hand, when the classical dynamics is chaotic, quantum fluctuations grow exponentially fast as $e^{\kappa t}/N^{1/2}$ with a positive constant $\kappa$, which is related to the Kolmogorov-Sinai entropy, and hence the relaxation time scales as $\ln N$.
In a spin-1/2 system, the classical dynamics is always regular since the equal-energy surface is one dimensional, and thus the relaxation time always scales as $N^{1/2}$.
In other systems such as a certain spin-1 system~\cite{Mori2017_classical}, the problem is reduced to the classical dynamics with a few degrees of freedom in the thermodynamic limit, and the classical phase space is divided into the regular region and the chaotic region.
The relaxation time scales as $N^{1/2}$ when the classical dynamics is regular, and as $\ln N$ when the classical dynamics is chaotic.

For $\alpha\neq 0$, $\{\hat{P}_{ij}\}$ are no longer conserved, and we cannot restrict ourselves into the TSS.
In this case, we should consider the DTWA, which reduces the problem to the classical dynamics of \textit{many} degrees of freedom starting from the initial distribution function (the discrete Wigner function) on the \textit{discrete} phase space.
We will see that for $\alpha\neq 0$ and $h_x\neq 0$, the classical dynamics becomes chaotic, and as a result, the relaxation time towards a prethermal state is always proportional to $\ln N$.
The second relaxation from a prethermal state is triggered by the relaxation of $\{\hat{P}_{ij}\}$ in a timescale not less than $\mathcal{O}(N^{1-\alpha})$ as we will see in Sec.~\ref{sec:proof}.

\subsection{$h_x=0$}
\label{sec:h0}

When $h_x=0$, $\hat{\sigma}_i^z$ is conserved for every $i$, and we can calculate the time evolution of spin operators exactly.
Let us define $\hat{\sigma}_i^{\pm}:=(\hat{\sigma}_i^x\pm i\hat{\sigma}_i^y)/2$.
In the Heisenberg picture, by using the formula $f(\hat{\sigma}^z)\hat{\sigma}^{\pm}=\hat{\sigma}^{\pm}f(\hat{\sigma}^z\pm 2)$, $\hat{\sigma}_i^{\pm}(t)=e^{i\hat{H}t}\hat{\sigma}_i^{\pm}e^{-i\hat{H}t}$ is calculated as
\beq
\hat{\sigma}_i^{\pm}(t)=\hat{\sigma}_i^{\pm}e^{\mp 2i\left(\sum_{j(\neq i)}^NJ_{ij}\hat{\sigma}_j^z+h_z\right)t}
=\hat{\sigma}_i^{\pm}e^{\mp 2ih_zt}\prod_{j(\neq i)}^N\left[\cos\left(2J_{ij}t\right)\mp i\hat{\sigma}_j^z\sin\left(2J_{ij}t\right)\right].
\eeq
For a factorized initial state, the expectation value at time $t$ is calculated as
\beq
\<\hat{\sigma}_i^{\pm}\>_t=\<\hat{\sigma}_i^{\pm}\>_0e^{\mp 2ih_zt}\prod_{j(\neq i)}^N\left[\cos\left(2J_{ij}t\right)\mp i\<\hat{\sigma}_j^z\>_0\sin\left(2J_{ij}t\right)\right],
\eeq
where we used the following notation: $\<\hat{O}\>_t:=\<\Psi(t)|\hat{O}|\Psi(t)\>=\<\Psi(0)|\hat{O}(t)|\Psi(0)\>$.
In particular, for the initial state given by Eq.~(\ref{eq:initial}), $\<\hat{\sigma}_i^\pm\>_0=1/2$ and $\<\hat{\sigma}_j^z\>=0$, and thus we have
\beq
\<\hat{\sigma}_i^{\pm}\>_t=\frac{e^{\mp 2ih_zt}}{2}\prod_{j(\neq i)}^N\cos\left(2J_{ij}t\right).
\eeq
Since $J_{ij}\propto N^{-(1-\alpha)}\ll 1$ for each $(i,j)$, $\<\hat{\sigma}_i^{\pm}\>_t$ is approximated as
\beq
\<\hat{\sigma}_i^{\pm}\>_t\approx\frac{e^{\mp 2ih_zt}}{2}\prod_{j(\neq i)}^N\left(1-2J_{ij}^2t^2\right)
\approx \<\sigma_i^{\pm}\>_0e^{\mp 2ih_zt}\exp\left(-2\sum_{j(\neq i)}^NJ_{ij}^2t^2\right).
\label{eq:S+-}
\eeq
The sum $\sum_{j(\neq i)}^NJ_{ij}^2$ is evaluated as
\beq
\sum_{j(\neq i)}^NJ_{ij}^2=\frac{J_{\alpha}^2}{N^{2(1-\alpha)}}\sum_{j(\neq i)}\frac{1}{r_{ij}^{2\alpha}}
\propto\left\{
\begin{split}
N^{-1}& \quad\text{for } 0\leq\alpha<\frac{1}{2}, \\
N^{-1}\ln N& \quad\text{for } \alpha=\frac{1}{2}, \\
N^{-2(1-\alpha)}& \quad\text{for } \frac{1}{2}<\alpha<1.
\end{split}
\right.
\label{eq:J^2}
\eeq
From Eq.~(\ref{eq:S+-}), the relaxation time of $\<\hat{\sigma}_i^{\pm}\>_t$ is given by
\beq
\tau_{\mathrm{rel}}\sim\left(\sum_{j(\neq i)}^NJ_{ij}^2\right)^{-1/2}
\propto
\left\{
\begin{split}
N^{1/2}& \quad\text{for } 0\leq\alpha<\frac{1}{2}, \\
\left(\frac{N}{\ln N}\right)^{1/2}& \quad\text{for } \alpha=\frac{1}{2}, \\
N^{1-\alpha}& \quad\text{for } \frac{1}{2}<\alpha<1.
\end{split}
\right.
\label{eq:hx0_rel}
\eeq
Strictly speaking, the approximation made in Eq.~(\ref{eq:S+-}) is justified only when $t\ll N^{1-\alpha}$.
Therefore, one might think that the conclusion of $\tau_{\mathrm{rel}}\propto N^{1-\alpha}$ for $1/2<\alpha<1$ is not reliable.
Indeed, what one can conclude from the above calculation for $1/2<\alpha<1$ is that the relaxation time cannot be shorter than $N^{1-\alpha}$, i.e., a \textit{lower} bound of the relaxation time is proportional to $N^{1-\alpha}$.
On the other hand, Kastner~\cite{Kastner2011} derived a rigorous \textit{upper} bound of the relaxation time for $1/2<\alpha<1$ which is also proportional to $N^{1-\alpha}$.
Therefore, by combining them, one can safely conclude that $\tau_{\mathrm{rel}}\propto N^{1-\alpha}$ for $1/2<\alpha<1$.

As far as we look at the expectation values of single spin operators, there is no two-step relaxation.
It is shown by van den Worm et al.~\cite{Worm2013} that a two-step relaxation takes place for $0<\alpha<1/2$ if we look at the time evolution of a two-spin correlation function $\<\hat{\sigma}_i^x\hat{\sigma}_j^x\>_t$ in which the distance $r_{ij}$ is independent of the system size $N$.
By repeating similar calculations, we obtain
\beq
\<\hat{\sigma}_i^x\hat{\sigma}_j^x\>_t\simeq\frac{\cos(4h_zt)}{2}\exp\left(-2\sum_{k=1}^N(J_{ik}+J_{jk})^2t^2\right)+\frac{1}{2}\exp\left(-2\sum_{k=1}^N(J_{ik}-J_{jk})^2t^2\right),
\label{eq:hx0sxsx}
\eeq
where we put $J_{ii}=0$ for all $i=1,2,\dots, N$.
In the first term of the right hand side of Eq.~(\ref{eq:hx0sxsx}), the system size dependence of $\sum_{k=1}^N(J_{ik}+J_{jk})^2$ is identical to that of Eq.~(\ref{eq:J^2}), while in the second term,
\beq
\sum_{k=1}^N(J_{ik}-J_{jk})^2\sim\frac{1}{N^{2(1-\alpha)}}\int_0^Ndr\,\frac{1}{r^{2(\alpha+1)}}\sim N^{-2(1-\alpha)}
\eeq
for any $\alpha\in(0,1)$, and hence, the relaxation time of the second term of Eq.~(\ref{eq:hx0sxsx}) is always proportional to $N^{1-\alpha}$.
By comparing it with the relaxation time of the first term of Eq.~(\ref{eq:hx0sxsx}) given by Eq.~(\ref{eq:hx0_rel}), we see that the first term relaxes earlier than the second term for $0<\alpha\leq 1/2$, and prethermalization occurs.
In this case, the initial relaxation time is given by $\tau_{\mathrm{ini}}\propto N^{1/2}$ and the timescale of the second relaxation is given by $\tau_{\mathrm{rel}}\propto N^{1-\alpha}$.
For $1/2<\alpha<1$, there is no timescale separation and no prethermalization.
The relaxation time for $1/2<\alpha<1$ is proportional to $N^{1-\alpha}$.

\section{Main result}
\label{sec:main}

We consider a general situation with $0<\alpha<1$ and $h_x\neq 0$.
Our strategy is twofold.
First, we prove that every $\hat{P}_{ij}$ with $i$ and $j$ such that $r_{ij}$ is independent of $N$ is a quasi-conserved quantity with the relaxation time $\tau_{\mathrm{rel}}\propto N^{1-\alpha}$ by deriving the following inequality:
\beq
\left|\<\hat{P}_{ij}\>_t-\<\hat{P}_{ij}\>_0\right|\leq r_{ij}c_{\alpha}J_{\alpha}\frac{t}{N^{1-\alpha}},
\label{eq:theorem}
\eeq
where $c_{\alpha}:=8\alpha\sum_{n=1}^{\infty}(1/n^{\alpha+1})$, which is finite and independent of $N$.
Since $\<\hat{P}_{ij}\>_0=1$ in our initial state (\ref{eq:initial}), $\<\hat{P}_{ij}\>_t\approx 1$ up to $\tau_{\mathrm{rel}}\propto N^{1-\alpha}$ for any $i$ and $j$ with $r_{ij}$ not so large.
The system does not have the global permutation symmetry for $\alpha\neq 0$, but have the local permutation symmetry approximately.
Second, we evaluate the timescale of the initial relaxation by deriving and solving the equations of motion of spin-spin correlation functions $G_k^{ab}(t):=\<\delta\hat{\sigma}_1^a\delta\hat{\sigma}_k^b\>_t$ ($a,b=x,y,z$) with $\delta\hat{\bm{\sigma}}_i:=\hat{\bm{\sigma}}_i-\<\hat{\bm{\sigma}}_i\>_t$.
In deriving those equations of motion, we assumed that $\delta\hat{\bm{\sigma}}_i$ is very small, and hence we can neglect all the higher order correlations such as $\<\delta\hat{\sigma}_i^a\delta\hat{\sigma}_j^b\delta\hat{\sigma}_k^c\>_t$, $\<\delta\hat{\sigma}_i^a\delta\hat{\sigma}_j^b\delta \hat{\sigma}_k^c\delta\hat{\sigma}_l^d\>_t$, and so on.
This approximation is equivalent to the truncation of the Bogoliubov-Born-Green-Kirkwood-Yvon (BBGKY) hierarchy at second order, and is justified as far as the spin-spin correlations $G_k^{ab}(t)$ are small.
It is noted that, in the thermodynamic limit, the MF approximation is exact and $G_k^{ab}(t)=0$ forever.
The initial stage of relaxation must be associated with the growth of quantum correlations $G_k^{ab}(t)$, and hence the timescale of the initial relaxation $\tau_{\mathrm{ini}}$ can be evaluated as a time at which $G_k^{ab}(t)$ exceeds a certain value.
If $\tau_{\mathrm{ini}}\ll\tau_{\mathrm{rel}}\propto N^{1-\alpha}$, there is a big timescale separation and prethermalization should take place.

It turns out that $\tau_{\mathrm{ini}}\ll\tau_{\mathrm{rel}}$ holds for any $\alpha\in(0,1)$ when $h_x\neq 0$.
More precisely, for sufficiently large $N$, $\tau_{\mathrm{ini}}\propto \ln N$ for any $\alpha\in(0,1)$.
It is noted that there is a discrepancy between this result, $\tau_{\mathrm{ini}}\propto\ln N$, and the exact result at $h_x=0$, $\tau_{\mathrm{ini}}\propto N^\gamma$ with $\gamma=\min\{1/2,1-\alpha\}$.
Our DTWA calculations presented in Sec.~\ref{sec:DTWA} suggest that the chaoticity of many-body classical dynamics causes the logarithmic dependence $\tau_\mathrm{ini}\propto\ln N$ for nonzero $h_x$.
Therefore, the absence of prethermalization for $1/2<\alpha<1$ found in Refs.~\cite{Worm2013,Kastner2015} is not generally true; it should be considered to be an exceptional result due to the exact solvability at $h_x=0$.

\subsection{Quasi conservation of local permutation operators}
\label{sec:proof}

In this subsection, we prove Eq.~(\ref{eq:theorem}), which states that every local permutation operator, i.e., $\hat{P}_{ij}$ with $i$ and $j$ not far from each other, is a quasi-conserved quantity with the relaxation time not shorter than $\mathcal{O}(N^{1-\alpha})$.
The proof is almost straightforward.
The time derivative of $\<\hat{P}_{ij}\>_t=\<\Psi(t)|\hat{P}_{ij}|\Psi(t)\>$ is given by
\beq
\frac{d}{dt}\<\hat{P}_{ij}\>_t=-i\<[\hat{P}_{ij},\hat{H}]\>_t
=-i\<P_{ij}(\hat{H}-\hat{P}_{ij}\hat{H}\hat{P}_{ij})\>_t,
\eeq
where we used $\hat{P}_{ij}^2=1$.
Therefore,
\beq
\left|\<\hat{P}_{ij}\>_t-\<\hat{P}_{ij}\>_0\right|\leq t\|\hat{P}_{ij}(\hat{H}-\hat{P}_{ij}\hat{H}\hat{P}_{ij})\|\leq t\|\hat{H}-\hat{P}_{ij}\hat{H}\hat{P}_{ij}\|,
\label{eq:bound}
\eeq
where $\|\hat{O}\|:=\sup_{\Psi:\<\Psi|\Psi\>=1}\sqrt{\<\Psi|\hat{O}^{\dagger}\hat{O}|\Psi\>}$ is the operator norm of an operator $\hat{O}$.
Since $\hat{P}_{ij}\hat{H}\hat{P}_{ij}$ is the Hamiltonian in which $\hat{\bm{\sigma}}_i$ and $\hat{\bm{\sigma}}_j$ are interchanged,
\beq
\hat{H}-\hat{P}_{ij}\hat{H}\hat{P}_{ij}=-\sum_{k(\neq i,j)}^N(J_{ik}-J_{jk})(\hat{\sigma}_i^z-\hat{\sigma}_j^z)\sigma_k^z.
\eeq
By using $\|(\hat{\sigma}_i^z-\hat{\sigma}_j^z)\hat{\sigma}_k^z\|\leq 2$, we obtain
\beq
\|\hat{H}-\hat{P}_{ij}\hat{H}\hat{P}_{ij}\|\leq 2\sum_{k(\neq i,j)}^N|J_{ik}-J_{jk}|.
\eeq
We have
\begin{align}
|J_{ik}-J_{jk}|=\frac{J_{\alpha}}{N^{1-\alpha}}\left|\frac{1}{r_{ik}^{\alpha}}-\frac{1}{r_{jk}^{\alpha}}\right|
&\leq\frac{J_{\alpha}}{N^{1-\alpha}}\alpha |r_{ik}-r_{jk}|\max\left\{\frac{1}{r_{ik}^{\alpha+1}},\frac{1}{r_{jk}^{\alpha+1}}\right\}
\nonumber \\
&\leq\frac{J_{\alpha}}{N^{1-\alpha}}\alpha r_{ij}\left(\frac{1}{r_{ik}^{\alpha+1}}+\frac{1}{r_{jk}^{\alpha+1}}\right).
\end{align}
Since $\sum_{k(\neq i,j)}^N(1/r_{ik}^{\alpha+1})=\sum_{k(\neq i,j)}^N(1/r_{jk}^{\alpha+1})\leq 2\sum_{n=1}^{\infty}(1/n^{\alpha+1})=c_\alpha/(4\alpha)$, we obtain
\beq
\|\hat{H}-\hat{P}_{ij}\hat{H}\hat{P}_{ij}\|\leq r_{ij}c_{\alpha}J_{\alpha}\frac{1}{N^{1-\alpha}}.
\eeq
By substituting it into Eq.~(\ref{eq:bound}), we obtain Eq.~(\ref{eq:theorem}).

It is noted that Eq.~(\ref{eq:theorem}) is generally true; an analogous inequality is also derived for other interactions such as the Heisenberg interaction $\sum_{ij}J_{ij}\hat{\bm{\sigma}}_i\cdot\hat{\bm{\sigma}}_j$.
It is obvious from the above derivation that Eq.~(\ref{eq:theorem}) holds for any initial state, which is not necessarily of the product form given in Eq.~(\ref{eq:initial}), and for any boundary condition.

\subsection{Timescale of the initial stage of relaxation}
\label{sec:timescale}

In the thermodynamic limit, the time evolution of a single spin is exactly given by the MF equation~(\ref{eq:MF_Heisenberg}), and spin correlation functions $G_k^{ab}(t)=\<\delta\hat{\sigma}_1^a\delta\hat{\sigma}_k^b\>_t$ are zero for any fixed time $t$.
When $N$ is large but finite, spin correlations grow with time, which gives rise to deviation from the MF approximation and drives the early stage of relaxation.

We shall derive the equations of motion of single spin expectation values $\<\hat{\bm{\sigma}}_i\>_t$ and spin correlation functions $G_k^{ab}(t)$ up to the leading order corrections in $1/N$.
It is noted that $\<\hat{\bm{\sigma}}_i\>_t$ is independent of $i$ and $G_k^{ab}(t)=\<\delta\hat{\sigma}_1^a\delta\hat{\sigma}_k^b\>_t=\<\delta\hat{\sigma}_{i+1}^a\delta\hat{\sigma}_{i+k}^b\>_t$ because of the translation symmetry of the Hamiltonian.
Thus, we can write $\<\hat{\bm{\sigma}}_i\>_t=\<\hat{\bm{\sigma}}\>_t$.
We decompose a spin operator in the Heisenberg picture $\hat{\bm{\sigma}}_i(t)$ as $\hat{\bm{\sigma}}_i(t)=\<\hat{\bm{\sigma}}\>_t+\delta\hat{\bm{\sigma}}_i(t)$.
By definition, $\<\delta\hat{\bm{\sigma}}_i\>_t=\<\delta\hat{\bm{\sigma}}_i(t)\>_0=0$.
The exact equations of motion for $\<\hat{\bm{\sigma}}\>_t$ and $\delta\hat{\bm{\sigma}}_i(t)$ are given as follows:
\beq
\left\{
\begin{split}
&\frac{d}{dt}\<\hat{\sigma}^x\>_t=2\<\hat{\sigma}^y\>_t\<\hat{\sigma}^z\>_t+2h_z\<\hat{\sigma}^y\>+2\sum_{j=2}^NJ_{ij}G_j^{yz}, \\
&\frac{d}{dt}\<\hat{\sigma}^y\>_t=-2\<\hat{\sigma}^x\>_t\<\hat{\sigma}^z\>_t-2h_z\<\hat{\sigma}^x\>+2h_x\<\hat{\sigma}^z\>_t-2\sum_{j=2}^NJ_{ij}G_j^{xz}, \\
&\frac{d}{dt}\<\hat{\sigma}^z\>_t=-2h_x\<\hat{\sigma}^y\>_t,
\end{split}
\right.
\label{eq:S}
\eeq
and
\beq
\left\{
\begin{split}
\frac{d}{dt}\delta\hat{\sigma}_i^x(t)=&2\<\hat{\sigma}^z\>_t\delta\hat{\sigma}_i^y(t)+2h_z\delta\hat{\sigma}_i^y(t)
\\
&+2\<\hat{\sigma}^y\>_t\sum_{j(\neq i)}^NJ_{ij}\delta\hat{\sigma}^z_j(t)+2\sum_{j(\neq i)}^NJ_{ij}\delta\hat{\sigma}_i^y(t)\delta\hat{\sigma}_j^z(t), \\
\frac{d}{dt}\delta\hat{\sigma}_i^y(t)=&-2\<\hat{\sigma}^z\>_t\delta\hat{\sigma}_i^x(t)-2h_z\delta\hat{\sigma}_i^x(t)+2h_x\delta\hat{\sigma}_i^z(t)
\\
&-2\<\hat{\sigma}^x\>_t\sum_{j(\neq i)}^NJ_{ij}\delta\hat{\sigma}_j^z(t)-2\sum_{j(\neq i)}J_{ij}\delta\hat{\sigma}_i^x(t)\delta\hat{\sigma}_j^z(t), \\
\frac{d}{dt}\delta\hat{\sigma}_i^z(t)=&-2h_x\delta\hat{\sigma}_i^y(t).
\end{split}
\right.
\label{eq:G0}
\eeq
Equation~(\ref{eq:G0}) is written as
\beq
\frac{d}{dt}\delta\hat{\bm{\sigma}}_i(t)=W\delta\hat{\bm{\sigma}}_i(t)+\vec{\Delta}_i+\vec{\xi}_i,
\label{eq:delta}
\eeq
where
\begin{align}
&\delta\hat{\bm{\sigma}}_i(t)=\begin{pmatrix}\delta\hat{\sigma}_i^x(t)\\ \delta\hat{\sigma}_i^y(t) \\ \delta\hat{\sigma}_i^z(t)\end{pmatrix}\\
&W=\begin{pmatrix}0 & 2(\<\hat{\sigma}^z\>_t+h_z) & 0 \\ -2(\<\hat{\sigma}^z\>_t+h_z) & 0 & 2h_x \\ 0 & -2h_x & 0\end{pmatrix}\\
&\vec{\Delta}_i=2\sum_{j(\neq i)}^NJ_{ij}\delta\hat{\sigma}_j^z\begin{pmatrix}\<\hat{\sigma}^y\>_t \\ -\<\hat{\sigma}^x\>_t \\ 0\end{pmatrix},
\label{eq:vec}
\end{align}
and
\beq
\vec{\xi}_i=2\sum_{j(\neq i)}J_{ij}\begin{pmatrix}\delta\hat{\sigma}_i^y(t)\delta\hat{\sigma}_j^z(t) \\ -\delta \hat{\sigma}_i^x(t)\delta\hat{\sigma}_j^z(t) \\ 0\end{pmatrix}.
\eeq
Since $\vec{\xi}_i$ gives higher order corrections in $1/N$, we neglect it.
This approximation is equivalent to the truncation of the BBGKY hierarchy at the second order.
By using Eq.~(\ref{eq:delta}), the equations of motion for $G_k^{ab}(t)$ are obtained:
\begin{align}
&\frac{d}{dt}G_k^{ab}(t)=\left\<\frac{d\delta\hat{\sigma}_i^a(t)}{dt}\delta\hat{\sigma}_j^b(t)\right\>+\left\<\delta\hat{\sigma}_i^a(t)\frac{d\delta\hat{\sigma}_j^b(t)}{dt}\right>
\nonumber \\
&=\sum_{c=x,y,z}\left(W_{ac}G_k^{cb}+W_{bc}G_k^{ac}\right)+v_a\sum_{j(\neq 1,k)}J_{1j}G_{r_{jk}+1}^{zb}+v_b\sum_{j(\neq 1,k)}J_{kj}G_j^{az}+J_{1k}(v_af_b+v_bf_a),
\label{eq:G}
\end{align}
where $\vec{v}$ is defined by $\vec{\Delta}_i=\sum_{j(\neq i)}^NJ_{ij}\delta\hat{\sigma}_j^z(t)\vec{v}$, see Eq.~(\ref{eq:vec}), and
\beq
\vec{f}=\begin{pmatrix}-\<\hat{\sigma}^x\>_t\<\hat{\sigma}^z\>_t \\ -\<\hat{\sigma}^y\>_t\<\hat{\sigma}^z\>_t \\ 1-\<\hat{\sigma}^z\>_t^2\end{pmatrix}.
\eeq
Equations (\ref{eq:S}) and (\ref{eq:G}) are desired equations of motion with the leading order corrections in $1/N$.

In order to access larger system sizes, we perform a further approximation explained below.
Because of long-range nature of the interactions, it is expected that $G_k^{ab}(t)$ is a smooth and slowly varying function of $k$.
For a certain large integer $\ell$, we consider the case of $N=M\ell$ with an integer $M$ for simplicity.
We partition all the sites $k=1,2,\dots N$ into $M$ groups $B_m$ ($m=1,2,\dots,M$) defined as
\beq
B_m:=\{(m-1)\ell+1,(m-1)\ell+2,\dots,m\ell\}.
\eeq
Each group contains $\ell$ sites.
We assume that if $k$ and $k'$ belong to the same group, $G_k^{ab}(t)\approx G_{k'}^{ab}(t)$.
This approximation means that $G_k^{ab}(t)$ is approximated by some function $\tilde{G}_m^{ab}(t)$ for all $k\in B_m$.
We choose $M$ as $1\ll M\ll N$, and then this approximation drastically decreases the number of independent variables.
In this paper, we set $M=100$.
The validity of this approximation has been checked by confirming that the result is converged with respect to changing the value of $M$.

After this procedure replacing $G_k^{ab}(t)$ by $\tilde{G}_m^{ab}(t)$, we evaluate $\tau_{\mathrm{ini}}$ as the time at which $\max_{a,b}|\tilde{G}_1^{ab}(t)|$ exceeds some value $G^*$, say $G^*=0.2$.

\begin{figure}[t]
\begin{center}
\begin{minipage}[t]{0.48\hsize}
\includegraphics[width=7cm]{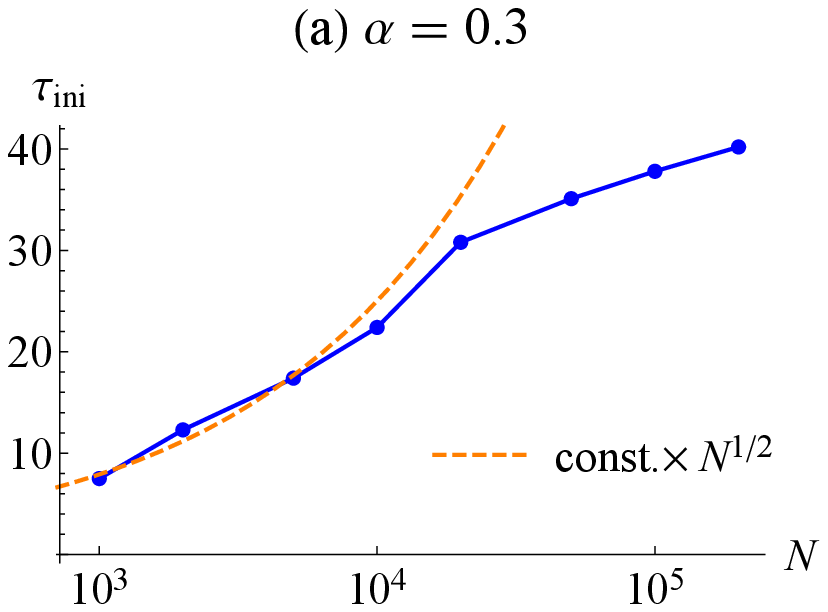}
\end{minipage}
\hfill
\begin{minipage}[t]{0.48\hsize}
\includegraphics[width=7.2cm]{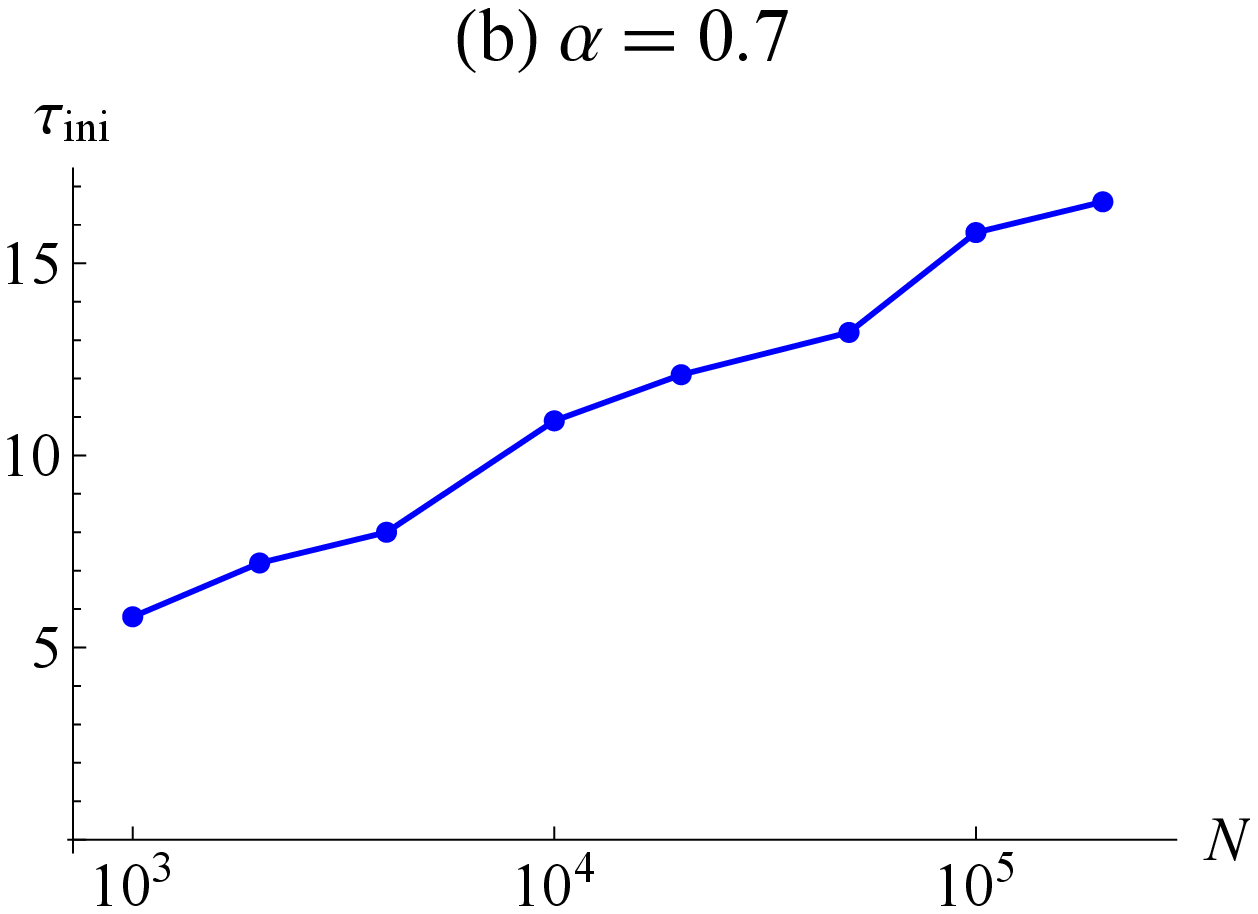}
\end{minipage}
\end{center}
\caption{(a) A semi-log plot of $\tau_{\mathrm{ini}}$ as a function of $N$ for $\alpha=0.3$.
We set $G^*=0.2$.
This graph shows $\tau_{\mathrm{ini}}\propto N^{1/2}$ for small $N$ but $\tau_\mathrm{ini}\propto \ln N$ for large $N$.
(b) A semi-log plot of $\tau_{\mathrm{ini}}$ as a function of $N$ for $\alpha=0.7$.
We set $G^*=0.2$.
This graph shows $\tau_{\mathrm{ini}}\propto \ln N$.}
\label{fig:tau_N}
\end{figure}

By numerically solving Eqs.~(\ref{eq:S}) and (\ref{eq:G}) for several different values of $N$ under the ansatz $G_k^{ab}(t)\approx\tilde{G}_m^{ab}(t)$ for all $k\in B_m$, we obtain the system-size dependence of $\tau_{\mathrm{ini}}$.
The parameters are set as $h_x=0.32$ and $h_z=0.26$.
In Fig.~\ref{fig:tau_N} (a), the result for $\alpha=0.3$ is shown.
We find $\tau_{\mathrm{ini}}\propto N^{1/2}$ for small $N$, but $\tau_\mathrm{ini}\propto\ln N$ for larger $N$ ($N\gtrsim 10^4$).
In Fig.~\ref{fig:tau_N} (b), the result for $\alpha=0.7$ is shown.
In this case, it is found that $\tau_{\mathrm{ini}}\propto\ln N$ even for $N\lesssim 10^4$.
Similar calculations for varying $\alpha$ suggest that $\tau_{\mathrm{ini}}\propto\ln N$ for sufficiently large $N$.
For smaller $\alpha$, the finite-size effect is stronger, and there is a crossover between $\tau_\mathrm{ini}\propto N^{1/2}$ for small $N$ and $\tau_\mathrm{ini}\propto\ln N$ for large $N$.

The behavior of $\tau_{\mathrm{ini}}\propto\ln N$ implies an exponentially fast growth of quantum correlations, which is the one expected for the chaotic classical dynamics (see section~\ref{sec:alpha0}) although the classical dynamics within the TSS is always regular.
In Sec.~\ref{sec:DTWA}, we will see that the semiclassical analysis using the DTWA indicates that the behavior $\tau_\mathrm{ini}\propto\ln N$ is related to the chaoticity of \textit{many-body} classical dynamics.

\section{Entire relaxation dynamics through the discrete truncated Wigner approximation}
\label{sec:DTWA}

Up to now, we have argued that (i) the relaxation of local permutation operators $\{\hat{P}_{ij}\}$ takes place in a timescale at least proportional to $N^{1-\alpha}$ and (ii) the initial relaxation time is proportional to $\ln N$ for sufficiently large $N$.
The statement (i) is mathematically rigorous, while (ii) is a numerical observation via an approximate treatment, i.e., the second-order truncation of the BBGKY hierarchy.

In this section, we use the DTWA to calculate the relaxation dynamics of several quantities.
As a result, it turns out that the initial relaxation time is indeed proportional to $\ln N$, and thus the statement (ii) is also verified by DTWA calculations.
Moreover, the DTWA enables us to calculate the entire relaxation dynamics, and we will see that a clear prethermalization plateau appears in the time evolution of spin-spin correlation functions.

\subsection{Discrete truncated Wigner approximation}
\label{sec:DTWA_review}

Let us begin with a brief review of the method of the DTWA.
Before discussing this method, let us first explain an well-established semiclassical method called the truncated Wigner approximation (TWA) for a quantum system with continuous canonical variables $(q,p)$~\cite{Polkovnikov2010}.
The DTWA is regarded as an extension of the TWA to systems with discrete degrees of freedom such as spin-1/2 systems.

For a given quantum state $|\psi\>$, we can introduce the Wigner function $W(q,p)$, which is a real-valued function on the phase space $(q,p)$.
Let us define the phase-point operator $\hat{A}(q,p)$ as
\beq
\<q'|\hat{A}(q,p)|q''\>=\delta\left(q-\frac{q'+q''}{2}\right)e^{ip(q'-q'')}.
\eeq
The Wigner function is then defined by
\beq
W(q,p)=\frac{1}{2\pi}\<\psi|\hat{A}(q,p)|\psi\>,
\eeq
or equivalently,
\beq
|\psi\>\<\psi|=\int dqdp\,W(q,p)\hat{A}(q,p).
\eeq
In general, $W(q,p)$ may not be nonnegative everywhere, and thus we cannot interpret $W(q,p)$ as the joint probability distribution of $q$ and $p$.
However, if we integrate $W(q,p)$ along a line $aq+bp=c$ with some real constants $a$, $b$, and $c$, it gives the probability density that the measurement outcome of the observable $a\hat{q}+b\hat{p}$ is given by $c$.
As special cases, if we set $b=0$, $P(q)=\int dp\,W(q,p)$ is the probability density of $q$, and if we set $a=0$, $P(p)=\int dq\, W(q,p)$ is the probability density of $p$.
The expectation value of an observable $O(\hat{q},\hat{p})$ is expressed as
\beq
\<\psi|O(\hat{q},\hat{p})|\psi\>=\int dqdp\, W(q,p)O_W(q,p),
\eeq
where the function $O_W(q,p)$ is called the Weyl symbol, which is defined by
\beq
O_W(q,p)=\Tr[O(\hat{q},\hat{p})\hat{A}(q,p)]
\eeq
After the time evolution, the expectation value is given by
\beq
\<\psi(t)|O(\hat{q},\hat{p})|\psi(t)\>=\int dqdp\,W(q,p)O_W(q,p;t),
\eeq
with
\beq
O_W(q,p;t)=\Tr[O(\hat{q},\hat{p})e^{-i\hat{H}t}\hat{A}(q,p)e^{i\hat{H}t}].
\eeq
The TWA replaces the quantum time evolution of the Weyl symbol by the classical one,
\beq
O_W(q,p;t)\approx O_W(q(t),p(t)),
\eeq
where $q(t)$ and $p(t)$ are the solutions of the classical equations of motion under the classical Hamiltonian $H(q,p)$,
\beq
\frac{dq(t)}{dt}=\frac{\d H(q,p)}{\d p}, \qquad \frac{dp(t)}{dt}=-\frac{\d H(q,p)}{\d q}.
\eeq
In the TWA, the effect of quantum fluctuations is incorporated only through the Wigner function of the initial state.
If the Wigner function $W(q,p)$ corresponding to the initial state is nonnegative everywhere, it can be regarded as the distribution function of the initial state.
In this case, what to do in TWA calculations is to statistically sample initial states and compute the classical time evolutions.

Next, we briefly explain the DTWA~\cite{Schachenmayer2015} for spin-1/2 systems.
For spin-1/2 systems, spin variables $\{(s_i^x,s_i^y,s_i^z)\}$, where $s_i^a=\pm 1$ is an eigenvalue of $\hat{\sigma}_i^a$ ($a=x,y,z$), are not continuous, but we can introduce the discrete phase space and the discrete Wigner function that has a similar property as the usual Wigner function~\cite{Wootters1987}.
For simplicity, let us consider a single spin-1/2.
The discrete phase space consists of the four points $(s^x,s^z)$ with $s^x=\pm 1$ and $s^z=\pm 1$.
We denote by $p$ one of the four phase-space points.
The assignment of the value of $s^y$ to each phase-space point is not unique.
The following two choices are familiar:
\beq
s^y=\begin{cases}
+1 &\text{if }s^x=s^z, \\
-1 &\text{otherwise},
\end{cases}
\label{eq:phase1}
\eeq
and
\beq
s^y=\begin{cases}
-1 &\text{if }s^x=s^z, \\
+1 &\text{otherwise}.
\end{cases}
\label{eq:phase2}
\eeq
We call the discrete phase space with the choice of Eq.~(\ref{eq:phase1}) the ``phase space I'' and that with the choice of Eq.~(\ref{eq:phase2}) the ``phase space II''.
In each case, spin variables $(s_p^x,s_p^y,s_p^z)$ are assigned to each point $p$ in the discrete phase space.

The discrete Wigner function $W_p$ for a state $|\psi\>$ is defined through the phase-point operator $\hat{A}(s^x,s^y,s^z)$ as
\beq
W_p=\frac{1}{2}\<\psi|\hat{A}(s_p^x,s_p^y,s_p^z)|\psi\>,
\eeq
where
\beq
\hat{A}(s^x,s^y,s^z)=\frac{1}{2}(1+s^x\hat{\sigma}^x+s^y\hat{\sigma}^y+s^z\hat{\sigma}^z).
\eeq
The discrete Wigner function may not be nonnegative everywhere, and thus it cannot be interpreted as the joint probability distribution of $(s^x,s^y,s^z)$.
However, if we sum up $W_p$ along a ``line'' in the discrete phase space, it gives a probability distribution, i.e.,
\beq
\left\{
\begin{split}
&\sum_{p:s_p^x=s^x}W_p=\mathrm{Prob}[\hat{\sigma}^x=s^x], \\
&\sum_{p:s_p^y=s^y}W_p=\mathrm{Prob}[\hat{\sigma}^y=s^y], \\
&\sum_{p:s_p^z=s^z}W_p=\mathrm{Prob}[\hat{\sigma}^z=s^z].
\end{split}
\right.
\eeq
The expectation value of an operator $\hat{O}$ is expressed as
\beq
\<\psi|\hat{O}|\psi\>=\sum_pW_pO_W(s_p^x,s_p^y,s_p^z),
\eeq
where the Weyl symbol $O_W$ is defined by
\beq
O_W(s^x,s^y,s^z)=\Tr[\hat{O}\hat{A}(s^x,s^y,s^z)].
\eeq

When $|\psi\>=|+_x\>$, $W_p=1/2$ for two phase-space points $p$ with $s_p^x=1$ and $W_p=0$ otherwise.
It means that $(s^x,s^y,s^z)=(1,1,1)$ or $(1,-1,-1)$ each with probability $1/2$ in the phase space I, and $(s^x,s^y,s^z)=(1,-1,1)$ or $(1,1,-1)$ each with probability $1/2$ in the phase space II.
Following the previous works~\cite{Schachenmayer2015, Pucci2016}, we mix up the two phase spaces; $(s^x,s^y,s^z)=(1,1,1), (1,1,-1), (1,-1,1)$, or $(1,-1,-1)$ each with probability $1/4$.
Following this probability distribution, the initial values of $(s^x,s^y,s^z)$ are statistically sampled.

For an $N$-spin system, the discrete phase space is defined by the set of points $\vec{p}=(p_1,p_2,\dots,p_N)$, where $p_i$ denotes a phase-space point for $i$th spin.
The discrete Wigner function for an $N$-spin state $|\Psi\>$ is given by
\beq
W_{\vec{p}}=\frac{1}{2^N}\<\Psi|\hat{A}(\{s_{p_i}^x,s_{p_i}^y,s_{p_i}^z\})|\Psi\>,
\eeq
where the phase-point operator for an $N$-spin system is given by
\beq
\hat{A}(\{s_i^x,s_i^y,s_i^z\})=\bigotimes_{i=1}^N\hat{A}(s_i^x,s_i^y,s_i^z).
\eeq

We now consider the time evolution.
The expectation value at time $t$ is given by
\beq
\<\Psi(t)|\hat{O}|\Psi(t)\>=\sum_{\vec{p}}W_{\vec{p}}\Tr\,\hat{O}e^{-i\hat{H}t}\left[\bigotimes_{i=1}^N\hat{A}(s_{p_i}^x,s_{p_i}^y,s_{p_i}^z)\right]e^{i\hat{H}t}.
\eeq
The quantum time evolution of the phase-point operator, $e^{-i\hat{H}t}[\bigotimes_{i=1}^N\hat{A}(s_{p_i}^x,s_{p_i}^y,s_{p_i}^z)]e^{i\hat{H}t}$, is complicated.
The DTWA corresponds to the repacement of this time evolution by the purely classical time evolution of spin variables:
\beq
e^{-i\hat{H}t}\left[\bigotimes_{i=1}^N\hat{A}(s_{p_i}^x,s_{p_i}^y,s_{p_i}^z)\right]e^{i\hat{H}t}\rightarrow\bigotimes_{i=1}^N\hat{A}(s_i^x(t,\vec{p}),s_i^y(t,\vec{p}),s_i^z(t,\vec{p})),
\eeq
where $\{\bm{s}_i(t,\vec{p})\}$ obey the classical equations of motion:
\beq
\frac{d}{dt}\bm{s}_i(t,\vec{p})=-2\bm{s}_i(t,\vec{p})\times\frac{\d H_\mathrm{cl}}{\partial\bm{s}_i}
\label{eq:classical_EOM}
\eeq
with the initial conditions $\bm{s}_i(0,\vec{p})=\bm{s}_{p_i}$ under the classical Hamiltonian $H_\mathrm{cl}$ corresponding to Eq.~(\ref{eq:H}):
\beq
H_\mathrm{cl}=-\sum_{i<j}^NJ_{ij}s_i^zs_j^z-h_z\sum_{i=1}^Ns_i^z-h_x\sum_{i=1}^Ns_i^x.
\eeq
In other words, the MF approximation is applied for the phase-point operator in the DTWA.

In summary, in the DTWA, the time evolution of a quantum spin system with the Hamiltonian~(\ref{eq:H}) starting from the initial state~(\ref{eq:initial}) is calculated by statistically sampling classical initial states $\{s_i^x=1,s_i^y,s_i^z\}$, where each of $s_i^y$ and $s_i^z$ is $+1$ or $-1$ with probability $1/2$, and solving the classical equations of motion~(\ref{eq:classical_EOM}).

\subsection{Comparison with the exact result for $h_x=0$}
\label{sec:DTWA_hx0}

\begin{figure}[t]
\begin{center}
\includegraphics[width=8cm]{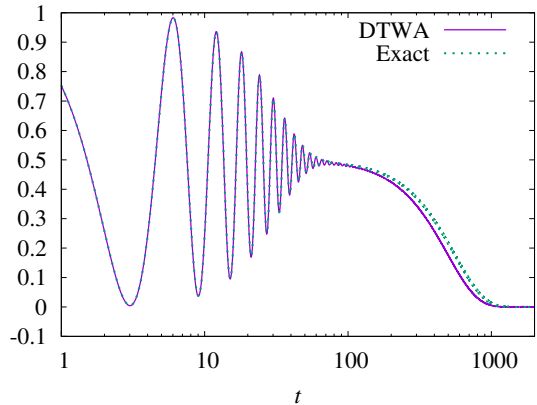}
\end{center}
\caption{Comparison between the time evolution obtained by the DTWA and the exact time evolution of the spin-spin correlation function $\<\hat{\sigma}_i^x\hat{\sigma}_{i+10}^x\>$.
In the DTWA simulation, we sampled 3456 initial states.
The agreement is fairly good for the entire relaxation process.}
\label{fig:compare}
\end{figure}

\begin{figure}[t]
\begin{center}
\begin{tabular}{c}
(a) $\alpha=0.3$ \\
\begin{minipage}[t]{0.49\hsize}
\includegraphics[width=7cm]{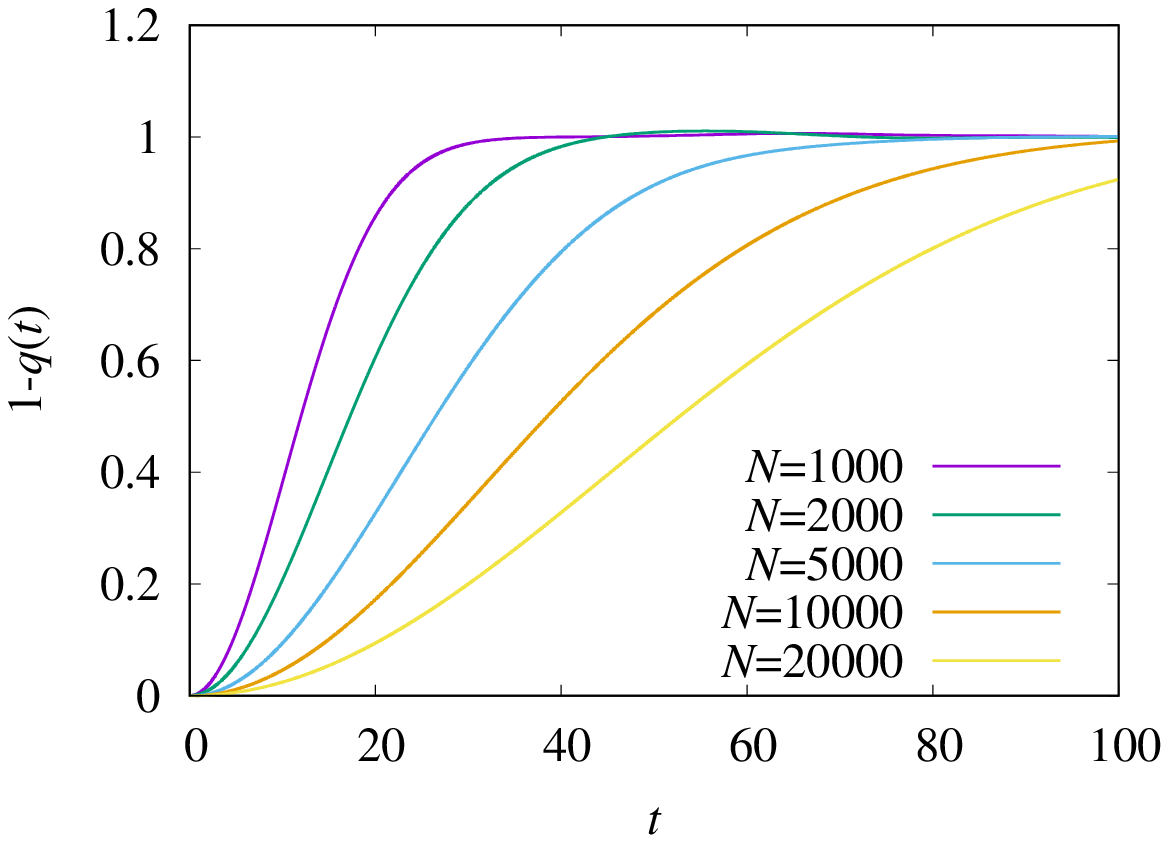}
\end{minipage}
\hfill
\begin{minipage}[t]{0.49\hsize}
\includegraphics[width=7cm]{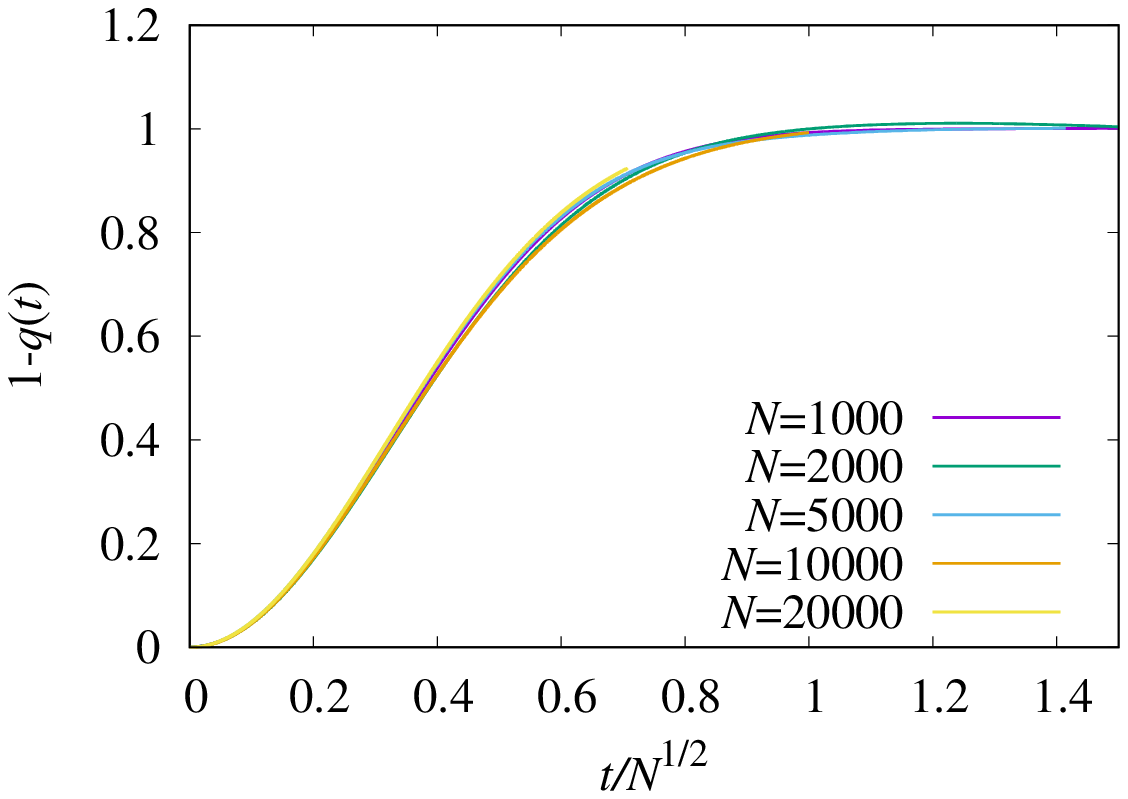}
\end{minipage}
\\
(b) $\alpha=0.7$ \\
\begin{minipage}[t]{0.49\hsize}
\includegraphics[width=7cm]{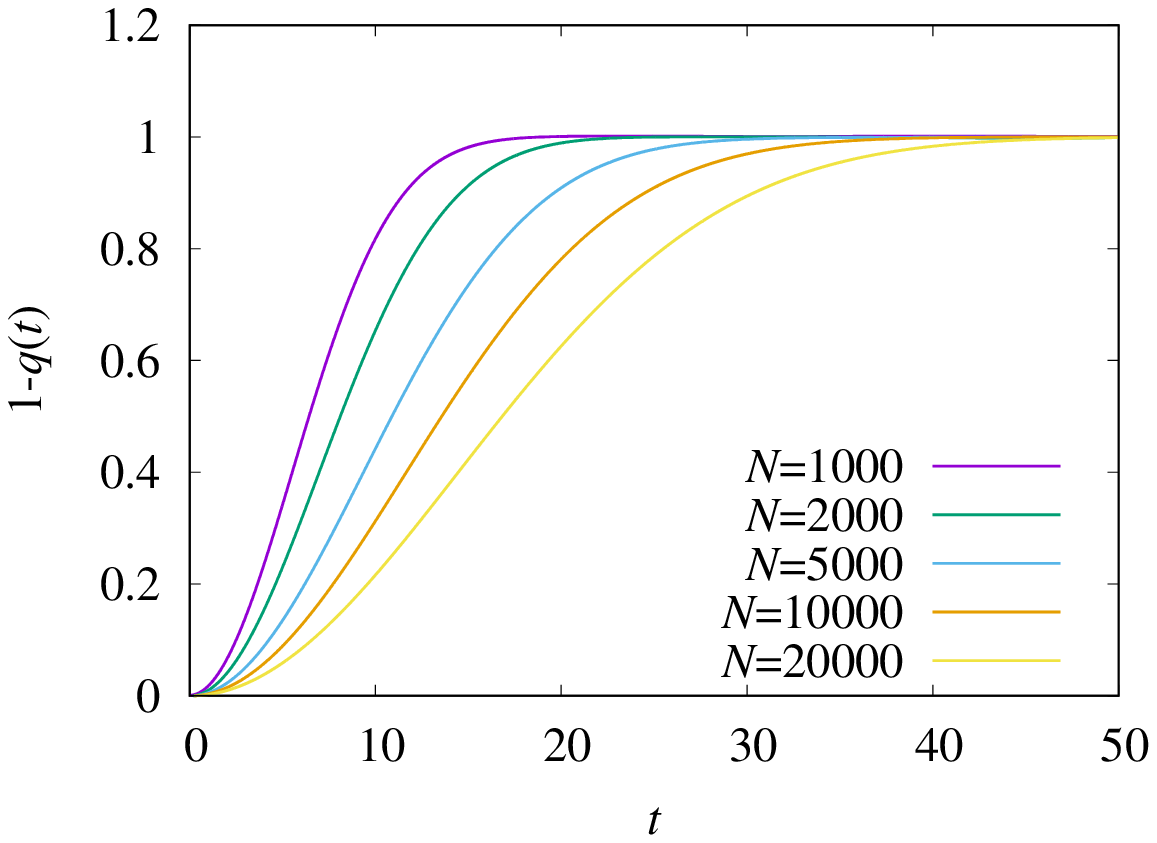}
\end{minipage}
\hfill
\begin{minipage}[t]{0.49\hsize}
\includegraphics[width=7cm]{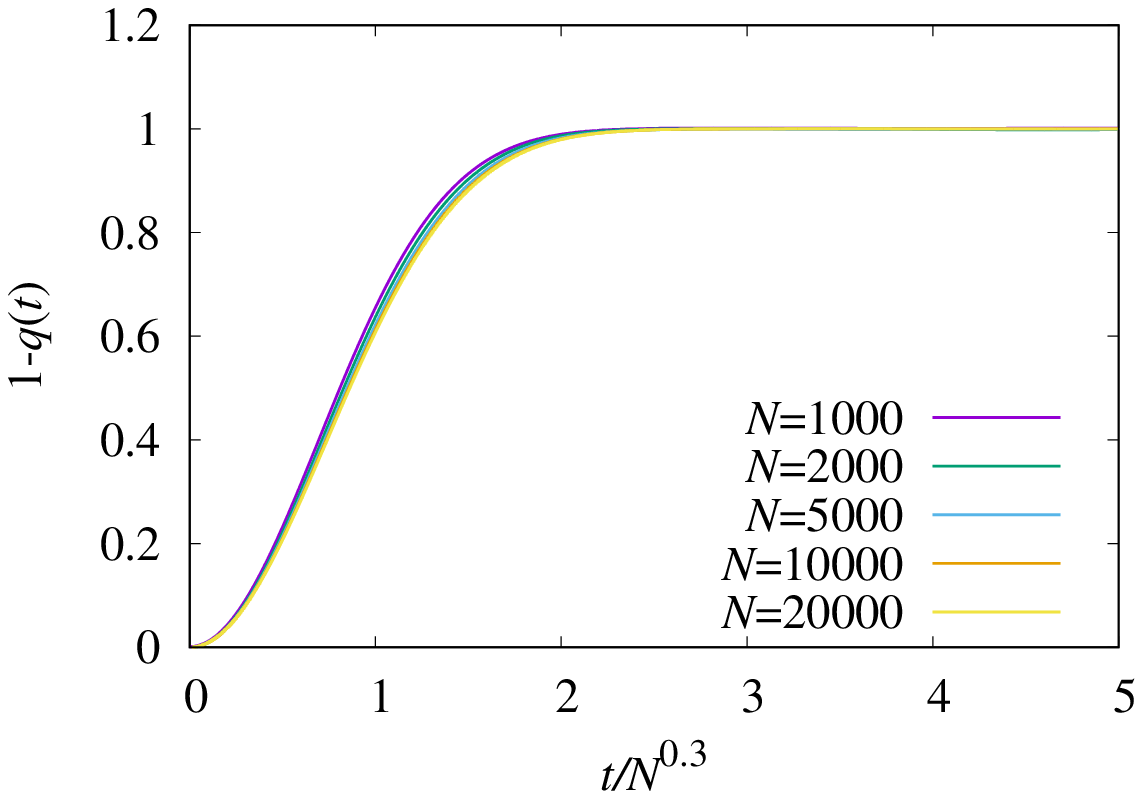}
\end{minipage}
\end{tabular}
\end{center}
\caption{The time evolution of the averaged overlap $q(t)$ for $h_x=0$, $h_z=0.26$.
In the right figure, the time axis is scaled as $t/N^{1/2}$ for (a) $\alpha=0,3$ and $t/N^{1-\alpha}=t/N^{0.3}$ for (b) $\alpha=0.7$.}
\label{fig:overlap_hx0}
\end{figure}

We first apply the DTWA to an exactly solvable case of $h_x=0$ and compare the result with the exact one presented in Sec.~\ref{sec:h0}.
We calculate the time evolution of a spin-spin correlation function $\<\hat{\sigma}_i^x\hat{\sigma}_{i+10}^x\>_t$, which is independent of $i$.
The magnetic field along $z$-direction is chosen as $h_z=0.26$.
In Fig.~\ref{fig:compare}, the DTWA result and the exact result are shown for $\alpha=0.3$.
Overall, the agreement between the DTWA result and the exact one is excellent.
In general, the DTWA reproduces an accurate result only for a short timescale, but the DTWA works very well in a long-range interacting system since the semiclassical approximation is expected to be good for such a system.
It should be however noted that the DTWA result does not completely converge to the exact one even in the thermodynamic limit.
The analytical relation between the DTWA result and the exact one for $h_x=0$ has been discussed in Ref.~\cite{Schachenmayer2015}.

We can estimate the timescale of the initial relaxation by introducing the averaged overlap $q(t)$.
First, we randomly choose two phase-space points $\vec{p}_1$ and $\vec{p}_2$ according to the discrete Wigner function $W_{\vec{p}}$.
After the classical time evolution starting from $\vec{p}_1$ and $\vec{p}_2$, we obtain two spin configurations $\{\bm{s}_i(t,\vec{p}_1)\}$ and $\{\bm{s}_i(t,\vec{p}_2)\}$, respectively, and calculate the overlap between them,
\beq
q_{12}(t)=\frac{1}{N}\sum_{i=1}^N\bm{s}_i(t,\vec{p}_1)\cdot\bm{s}_i(t,\vec{p}_2).
\eeq
We repeat this procedure many times, and calculate the averaged overlap $q(t)$:
\beq
q(t)=\sum_{\vec{p}_1}\sum_{\vec{p}_2}W_{\vec{p}_1}W_{\vec{p}_2}\frac{1}{N}\sum_{i=1}^N\bm{s}_i(t,\vec{p}_1)\cdot\bm{s}_i(t,\vec{p}_2).
\eeq
It is shown that $q(t)\in[-1,1]$, and the initial state of Eq.~(\ref{eq:initial}) gives $q(0)=1$.
After the time evolution, the averaged overlap $q(t)$ decays.
The speed of the decay of $q(t)$ is expected to give the timescale of the initial relaxation.
In Fig.~\ref{fig:overlap_hx0}, we show the time evolution of $1-q(t)$ in an exactly solvable case $h_x=0$ for (a) $\alpha=0.3$ and (b) $\alpha=0.7$.
If the time axis is scaled as $t/N^{1/2}$ for $\alpha=0.3$ and $t/N^{1-\alpha}=t/N^{0.3}$ for $\alpha=0.7$, the data for different system sizes are collapsed to a single curve.
From this fact, we can estimate $\tau_\mathrm{ini}$ as $\tau_\mathrm{ini}\propto N^{1/2}$ for $\alpha=0.3$ and $\tau_{\mathrm{ini}}\propto N^{1-\alpha}=N^{0.3}$ for $\alpha=0.7$, both of which are consistent with the exact results presented in Sec.~\ref{sec:h0}.

\subsection{Relaxation dynamics for $h_x\neq 0$}

\begin{figure}[t]
\begin{center}
\begin{minipage}[t]{0.49\hsize}
\includegraphics[width=7.5cm]{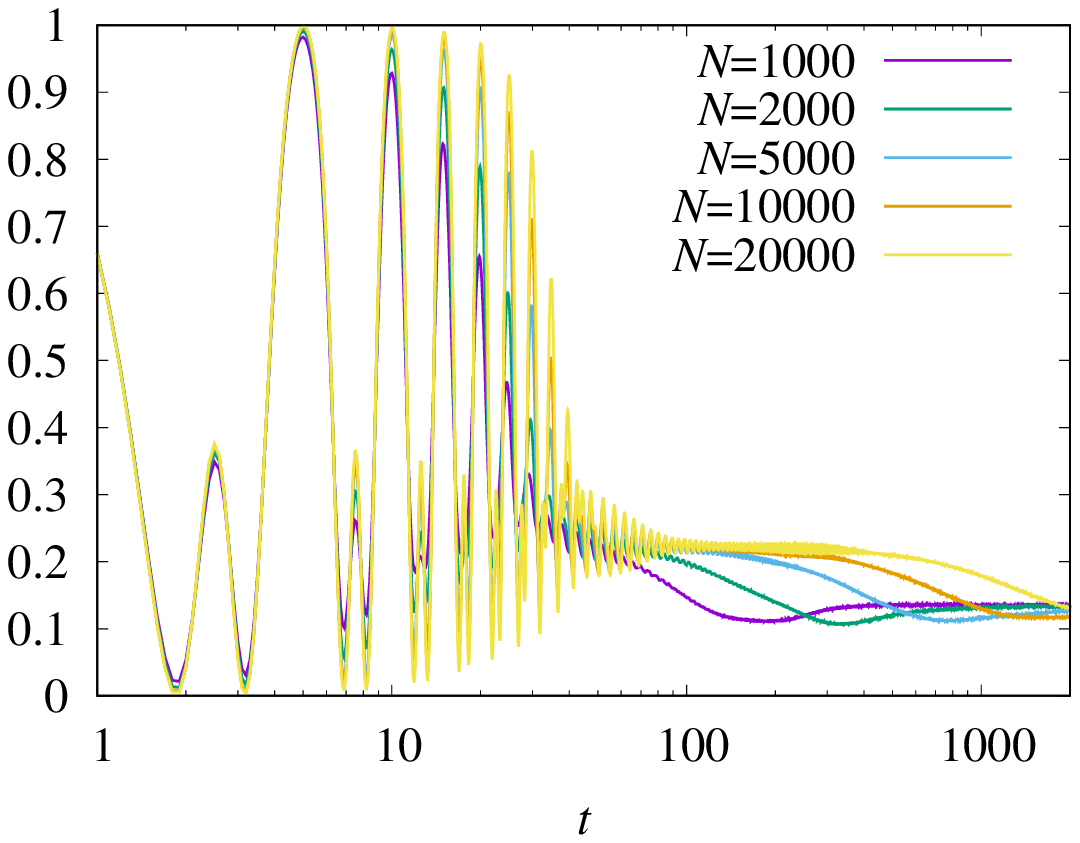}
\end{minipage}
\hfill
\begin{minipage}[t]{0.49\hsize}
\includegraphics[width=7.5cm]{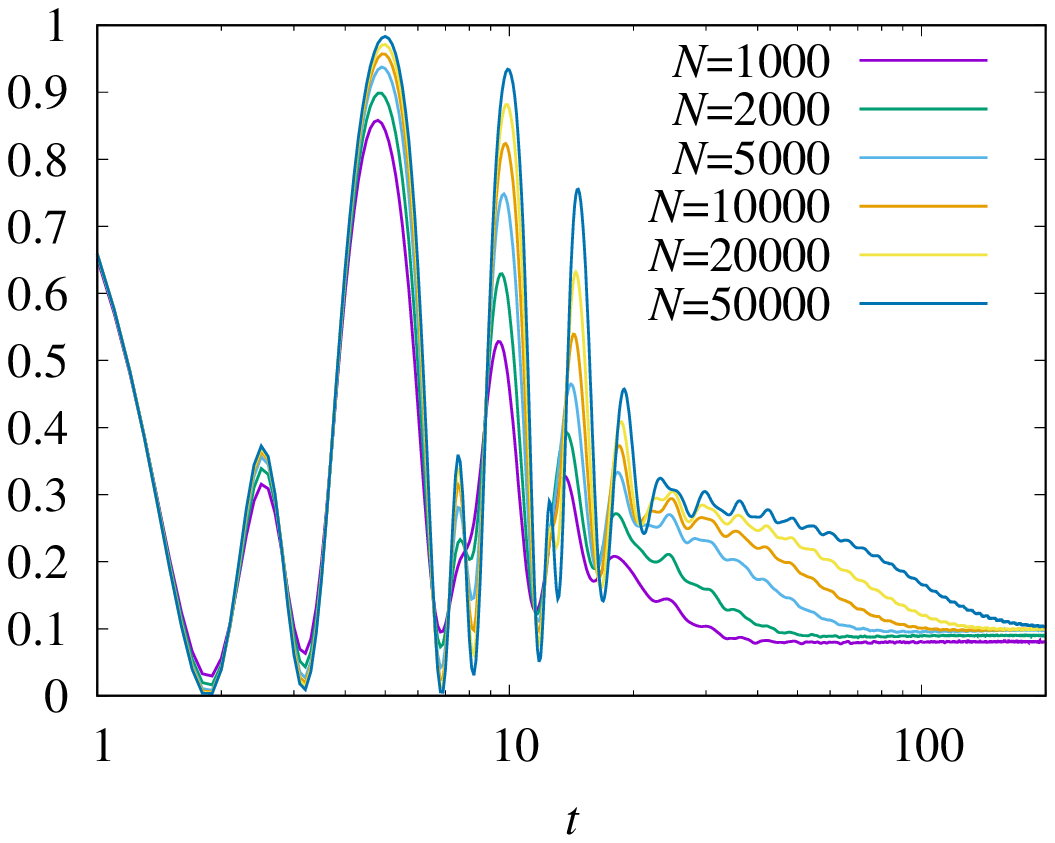}
\end{minipage}
\end{center}
\caption{Relaxation dynamics of $\<\hat{\sigma}_i^x\hat{\sigma}_{i+10}^x\>$ for (a) $\alpha=0.3$ and (b) $\alpha=0.7$.
The parameters are set as $h_x=0.32$ and $h_z=0.26$.}
\label{fig:DTWA}
\end{figure}

Now we apply the DTWA to the model with $h_x\neq 0$.
Here we set $h_x=0.32$ and $h_z=0.26$.
Figure~\ref{fig:DTWA} shows the time evolution of $\<\hat{\sigma}_i^x\hat{\sigma}_{i+10}^x\>_t$ calculated by the DTWA for (a) $\alpha=0.3$ and (b) $\alpha=0.7$.
We can see that prethermalization plateaus appear in both cases.

\begin{figure}[t]
\begin{center}
\begin{tabular}{c}
(a) $\alpha=0.3$ \\
\begin{minipage}[t]{0.49\hsize}
\includegraphics[width=7cm]{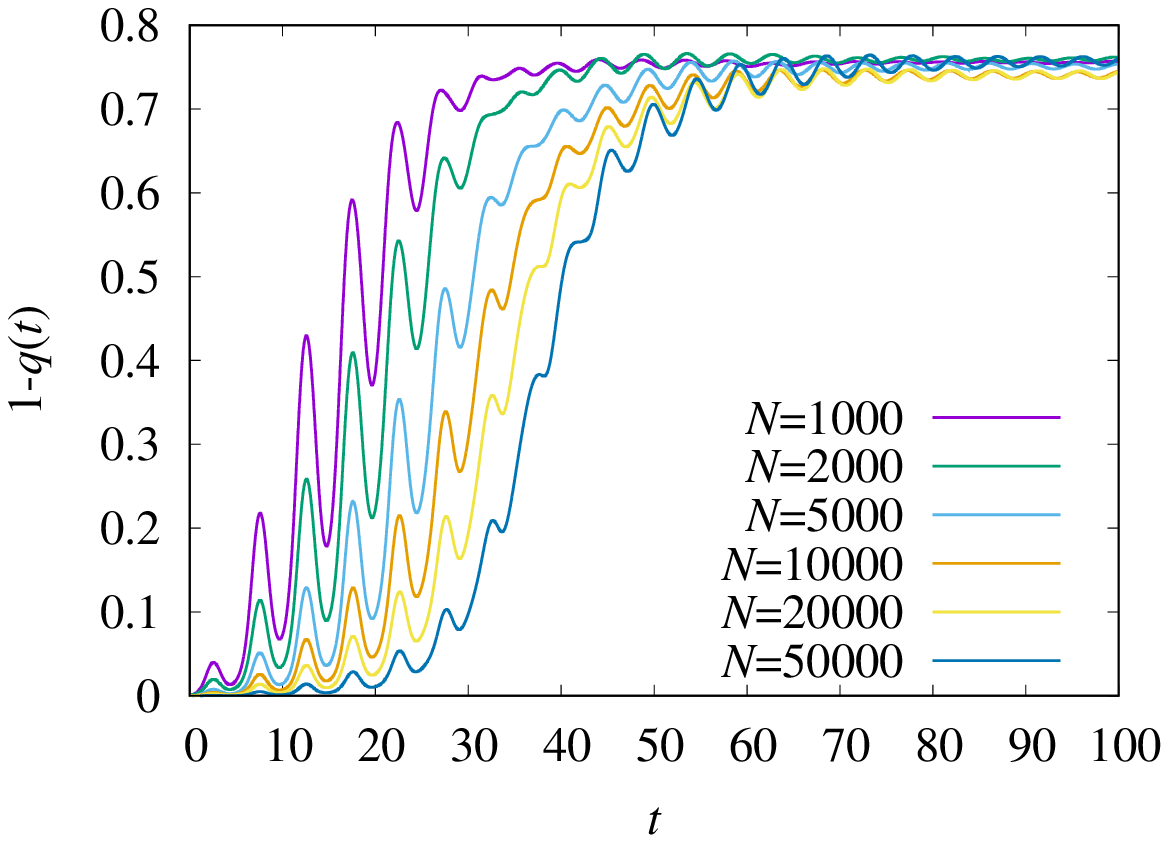}
\end{minipage}
\hfill
\begin{minipage}[t]{0.49\hsize}
\includegraphics[width=7cm]{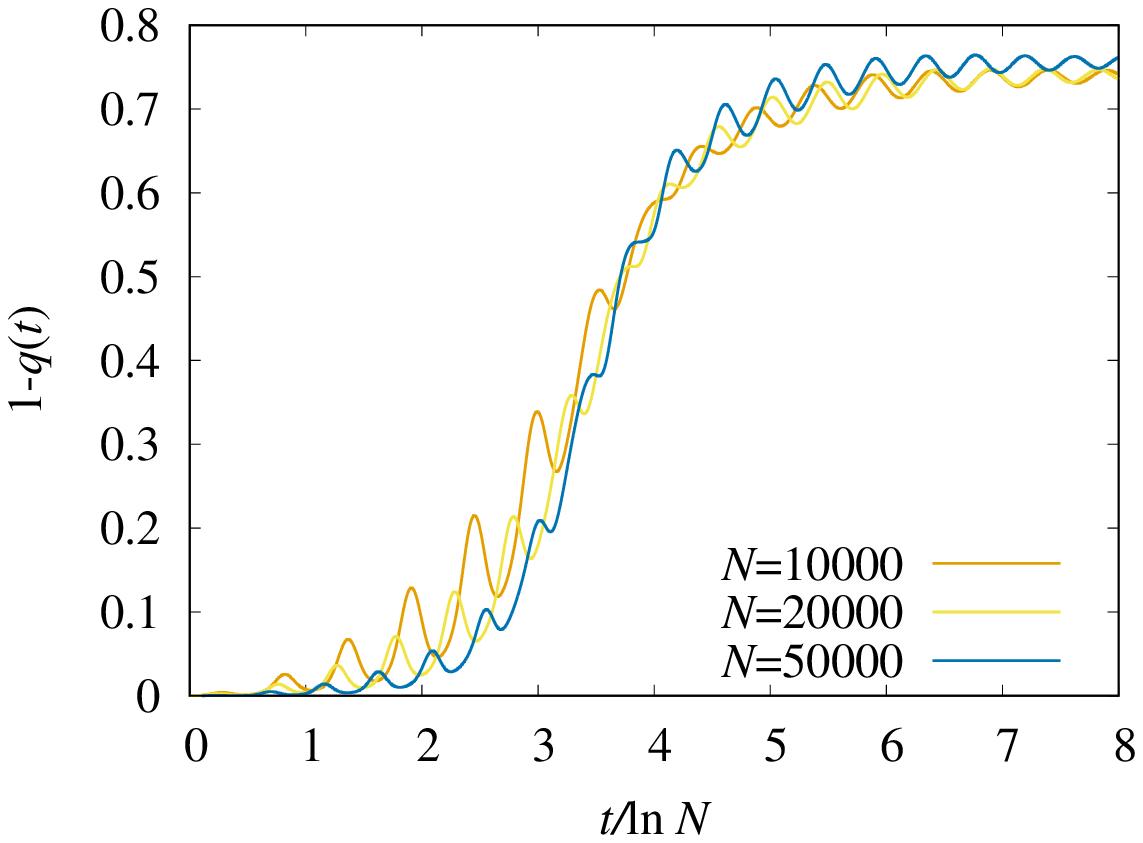}
\end{minipage}
\\
(b) $\alpha=0.7$ \\
\begin{minipage}[t]{0.49\hsize}
\includegraphics[width=7cm]{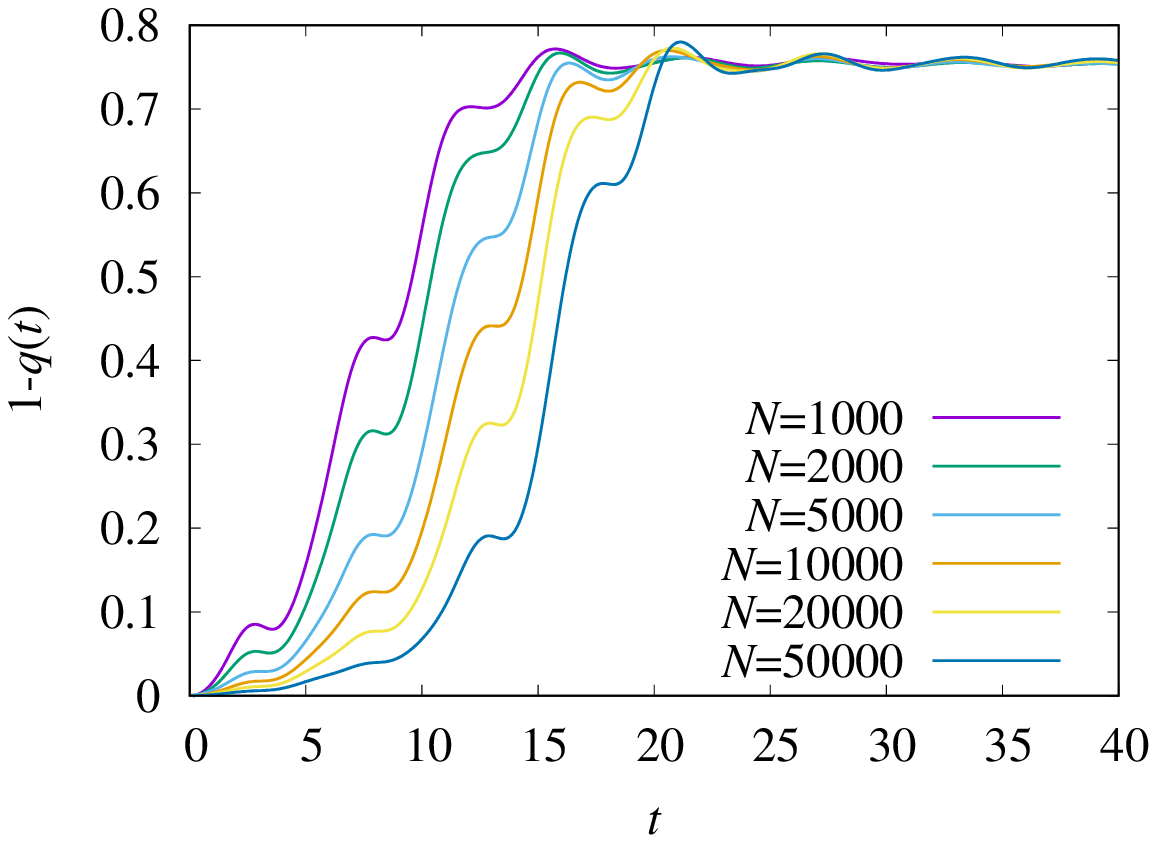}
\end{minipage}
\hfill
\begin{minipage}[t]{0.49\hsize}
\includegraphics[width=7cm]{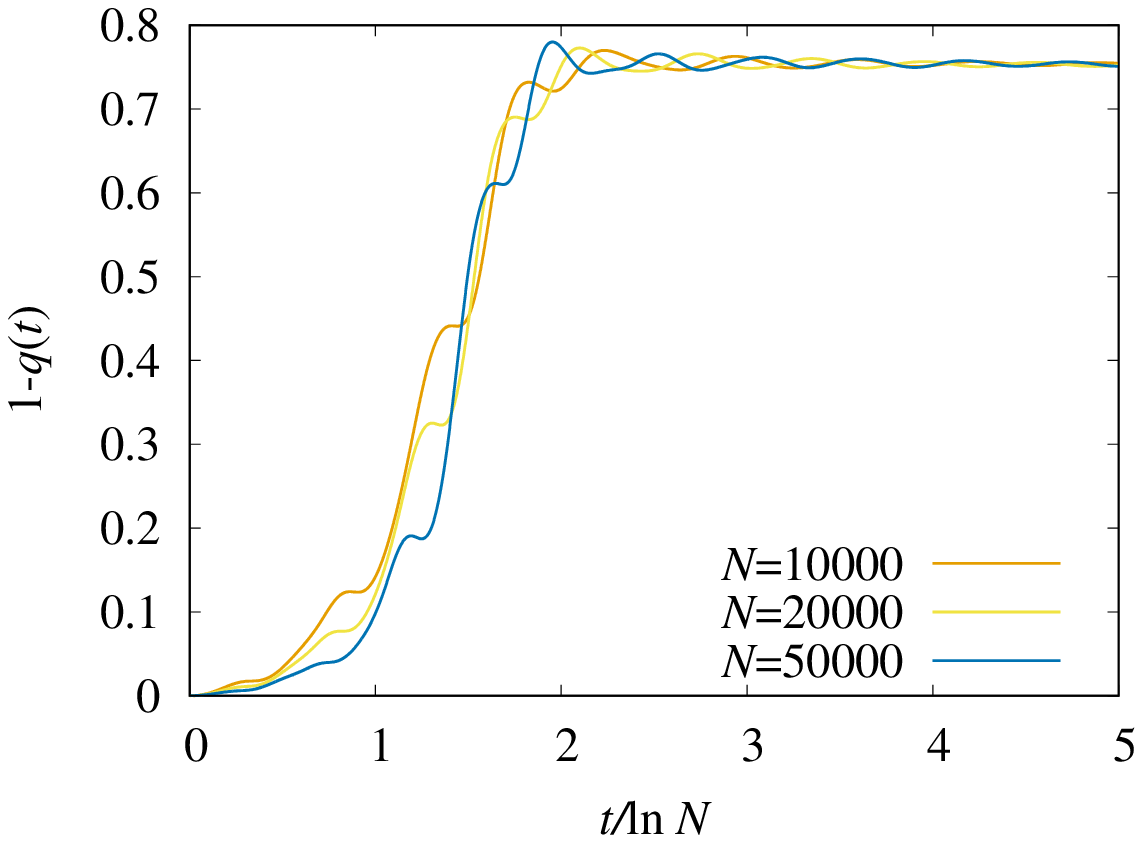}
\end{minipage}
\end{tabular}
\end{center}
\caption{The time evolution of averaged overlaps $q(t)$ for several system sizes with $h_x=0.32$, $h_z=0.26$.
In the right figure, the time axis is scaled as $t/\ln N$ both for (a) $\alpha=0,3$ and (b) $\alpha=0.7$.}
\label{fig:overlap}
\end{figure}

As already discussed in Sec.~\ref{sec:DTWA_hx0}, we can estimate the initial relaxation time by looking at the time evolution of the overlap $q(t)$.
Numerical results of the overlap dynamics for (a) $\alpha=0.3$ and (b) $\alpha=0.7$ are given in Fig.~\ref{fig:overlap}.
It turns out that the data are collapsed into a single curve for large $N$ ($N\geq 10^4$) by scaling the time axis as $t/\ln N$ for both $\alpha=0.3$ and 0.7.
This means that $\tau_\mathrm{ini}\propto\ln N$ for sufficiently large $N$.
Since the second relaxation occurs at a much longer timescale $\tau_\mathrm{rel}\propto N^{1-\alpha}$, there is a big timescale separation between $\tau_\mathrm{ini}$ and $\tau_\mathrm{rel}$, which results in prethermalization.

The behavior of $\tau_\mathrm{ini}\propto\ln N$ is similar to that for $\alpha=0$ when the underlying classical dynamics is chaotic (see Sec.\ref{sec:alpha0}).
It is expected that the overlap between two spin configurations rapidly decays in the chaotic classical dynamics, which explains why $\tau_\mathrm{ini}$ for $h_x\neq 0$ is much shorter than $\tau_\mathrm{ini}$ for $h_x=0$ (remember that the system is integrable and thus the underlying classical dynamics is not chaotic when $h_x=0$).
In this way, the result of $\tau_\mathrm{ini}\propto\ln N$ is understood as a consequence of the chaoticity of the classical dynamics.

A remaining question to be addressed is whether a state after the second relaxation is thermal equilibrium or not.
This is investigated by comparing the stationary value of $\<\hat{\sigma}_i^x\hat{\sigma}_{i+10}^x\>$ with its equilibrium value evaluated at the temperature corresponding to the initial value of the energy.
We can evaluate the equilibrium value by using the \textit{exactness of the mean-field theory} in an equilibrium state of a long-range interacting system, which is valid in classical~\cite{Cannas2000,Campa2000,Barre2001,Mori2013_phase} as well as quantum spin systems~\cite{Mori2012_equilibrium}.
According to Ref.~\cite{Mori2012_equilibrium}, when $\alpha\in(0,1)$, $\<\hat{\sigma}_i^x\hat{\sigma}_j^x\>_\mathrm{eq} \approx\<\hat{\sigma}_i^x\>_\mathrm{eq}\<\hat{\sigma}_j^x\>_\mathrm{eq}=:(m^x)^2$.
Here, $m^x$ is given by 
\beq
m^x=-\frac{\d}{\d h_x}f(\beta,h_x,h_z)
\eeq
with $f(\beta,h_x,h_z)$ being the free energy density at the inverse temperature $\beta$,
\begin{align}
f(\beta,h_x,h_z)=\min_{m^x,m^y,m^z}&\left[-\frac{1}{2}(m^z)^2-h_zm^z-h_xm^x
\right.\nonumber \\
&\left.-\frac{1}{\beta}\left(-\frac{1+|\bm{m}|}{2}\ln\frac{1+|\bm{m}|}{2}-\frac{1-|\bm{m}|}{2}\ln\frac{1-|\bm{m}|}{2}\right)\right],
\end{align}
where $|\bm{m}|=\sqrt{(m^x)^2+(m^y)^2+(m^z)^2}$.
It should be noted that the equilibrium free energy is independent of the value of $\alpha$.

In the method summarized above, we obtain the equilibrium value $\<\sigma_i^x\sigma_{j}^x\>_\mathrm{eq}\approx 0.04$, which does not agree to the stationary value after the second relaxation in Fig.~\ref{fig:DTWA}.
In our calculations, we cannot judge whether this discrepancy is due to an unphysical error by the DTWA or the existence of a longer timescale in which the system thermalizes eventually.
 
\section{Conclusion}
\label{sec:conclusion}

We have studied nonequilibrium dynamics of a long-range interacting spin chain isolated from the environment.
Because of long-range interactions decaying as $1/r^{\alpha}$ with distance $r$, we find that local permutation operators are almost conserved up to the relaxation time $\tau_{\mathrm{rel}}\propto N^{1-\alpha}$.
We have compared it to the timescale of the initial relaxation due to the growth of quantum correlations among spins, which is denoted by $\tau_{\mathrm{ini}}$.

When $h_x=0$, there is a big timescale separation and prethermalization occurs only for $0<\alpha\leq1/2$~\cite{Worm2013,Kastner2015}.
On the other hand, when $h_x\neq 0$, it turns out that a big timescale separation exists for any $\alpha\in(0,1)$.
The initial relaxation time behaves as $\tau_\mathrm{ini}\propto\ln N$, which is confirmed by the second-order truncation of the BBGKY hierarchy as well as the DTWA.
We have argued that the logarithmic dependence of $\tau_\mathrm{ini}$ on $N$ is due to the chaoticity of the underlying classical dynamics.

In the present DTWA calculations, a state after the second relaxation is not thermal equilibrium.
It is a future problem to understand whether this is due to an unphysical error introduced by the DTWA or due to the existence of a much longer timescale in which the system thermalizes.

Finally, it is pointed out that there may be another kind of timescale separations in long-range interacting systems.
Indeed, long-range Ising interactions are available in ion-trap experiments~\cite{Porras2004,Kim2009,Britton2012,Islam2013,Neyenhuis_arXiv2016}, and a big timescale separation of different origin was experimentally reported in Ref.~\cite{Neyenhuis_arXiv2016}.
Nonequilibrium dynamics in long-range interacting systems should be further examined theoretically and experimentally.

\section*{Acknowledgements}
The author thanks Kazuya Fujimoto for useful discussion.
The computation in this work has been done using the facilities of the
Supercomputer Center, the Institute for Solid State Physics, the University
of Tokyo.

\bibliography{LR_pre.bib}

\end{document}